\documentclass{article}
\usepackage{geometry}
\usepackage{xcolor}
\usepackage{amsmath}
\usepackage{amssymb}
\usepackage{authblk}            
\usepackage{graphicx}           
\usepackage{multirow}
\usepackage{bm}
\usepackage{caption}
\usepackage{subcaption}
\usepackage{soul}
\usepackage{hyphenat}
\usepackage{lipsum}
\usepackage{mathtools}
\usepackage{tikz}
\usepackage{tikz-3dplot}
\usetikzlibrary{shapes,calc,positioning}
\usepackage{pgfplots}
\pgfplotsset{compat=1.16}

\usepackage[backend=biber, style=numeric-comp, sorting=none]{biblatex}
\title{Quantum computing for classical problems: Variational Quantum Eigensolver for activated processes}
\author[1]{P.Pravatto}
\author[1]{D.Castaldo}
\author[1]{F.Gallina}
\author[1,2]{B.Fresch}
\author[1,2,3]{S.Corni}
\author[1,2]{G.J.Moro}
\affil[1]{Università degli studi di Padova, Dipartimento di Scienze Chimiche, Via Marzolo 1 - 35131 Padova (Italy)}
\affil[2]{Padua Quantum Technologies Research Center, Università di Padova}
\affil[3]{Istituto Nanoscienze—CNR, via Campi 213/A, 41125 Modena (Italy)}

\date{}                     
\begin{document}

\twocolumn[
    \begin{@twocolumnfalse}
        \maketitle
        \begin{abstract} The theory of stochastic processes impacts both physical and social sciences. At the molecular scale, stochastic dynamics is ubiquitous because of thermal fluctuations. The Fokker-Plank-Smoluchowski equation models the time evolution of the probability density of selected degrees of freedom in the diffusive regime and it is therefore a workhorse of physical chemistry.  In this paper we report the development and implementation of a Variational Quantum Eigensolver procedure to solve the Fokker-Planck-Smoluchowski eigenvalue problem. We show that such an algorithm, typically adopted to address quantum chemistry problems, can be applied effectively to classical systems paving the way to new applications of quantum computers. We compute the conformational transition rate in a linear chain of rotors experiencing nearest-neighbour interaction. We provide a method to encode on the quantum computer the probability distribution for a given conformation of the chain and assess its scalability in terms of operations. Performance analysis on noisy quantum emulators and quantum devices (IBMQ Santiago) is provided for a small chain showing results in good agreement with the classical benchmark without further addition of any error mitigation technique.\\
        \end{abstract}
    \end{@twocolumnfalse}
]

\section{Introduction}
\label{Introduction}

Since the dawn of quantum mechanics the study of quantum phenomena has been one of the key focus of the scientific community. For many years quantum systems have been investigated with the sole purpose of deepening our understanding on their microscopic behaviour. In recent years, however, scientific and technological advancement opened the way to the direct manipulation and control of quantum systems with the aim of solving practical problems. This shift in paradigm, sometime referred as second quantum revolution~\cite{deutsch2020harnessing}, changed the way of looking at the quantum world moving it from an object of study to the status of a powerful new tool. Quantum computation is surely one of the key examples of these new quantum technologies and unearthing computational tasks for which the use of quantum resources can offer a significant increase in efficiency with respect to the best classical algorithmic counterpart is one of the key challenges of the field. Many efforts of the scientific community have been devoted towards such an alternative paradigm of computation leading to the development of several algorithms that heralded its disruptive potential~\cite{terhal2018quantum}.
Particularly, with regard to relevant breakthroughs in physics and chemistry, quantum simulation is considered among the first near-term applications of this field~\cite{georgescu2014quantum}. In the present era of Noisy Intermediate Quantum Devices (NISQ)~\cite{mcclean2016theory}, hybrid quantum algorithms represent a valuable tool to boost classical computational power using small quantum processors. In this framework, some of the more performing algorithms are the so called Variational Hybrid Algorithms (VHA) where the quantum device is tasked with the preparation of a parametrized trial state and the computation of a cost function which, in turn, is minimized by an external classical optimization routine~\cite{mcclean2016theory}. This method requires the quantum device to run circuits of reasonably short-depth that cope with the coherence time and error rate of modern NISQ devices~\cite{jurcevic2021demonstration}.\\

The literature about VHA is constantly growing both in terms of different implementations and in terms of their applications. These range from the study of the electronic structure of small molecules~\cite{McArdle2020, Kandala2017} and molecular vibrations~\cite{Ollitrault2020} to the simulation of condensed matter and high-energy physics~\cite{yoshioka2020variational, nachman2021quantum}. In contrast, less attention has been dedicated to the application of quantum computing in the context of stochastic simulations (i.e., simulations of systems characterized by variables with a random character, and thus by classical probability distributions). We believe that leveraging standard methods of stochastic calculus, while harnessing the potential of quantum computers, may lead to the development of new quantum tools in the field of complex system simulations. This can possibly pave the way to important advances in chemistry, biochemistry and all the other connected fields in which stochastic analysis plays a relevant role such as quantitative finance, epidemiology, and computational fluid dynamics~\cite{haven2002discussion, lin2014long, singh2016gaseous}. Here we propose a new VHA to solve the Fokker\hyp{}Planck\hyp{}Smoluchowski (FPS) eigenvalue problem based on the isomorphism existing between the FPS operator and the quantum Hamiltonian~\cite{miyazawa1989theory, elber2020molecular}. 
As an example, in this paper we will focus on the approximation of the kinetic rate constant for the conformational transition in linear chain molecules that is connected with the first non-vanishing eigenvalue of the FPS operator. Despite of its simplicity, this model has been used as theoretical starting point for the conformational analysis in polymers~\cite{helfand1978brownian, moro1991coupling, moro1992coupling} whose study has deep implications in many different areas~\cite{doi1988theory}.\\

The present contribution is organized as follows: Section~\ref{sec:Theory} outlines the theoretical basis of our approach. We give a general overview on how the FPS eigenvalue problem can be treated on a quantum computer and then we formalize the stochastic description of the conformational dynamics of a chain of rotors characterized by a bistable intramolecular potential energy surface.\\

This system is the case-study of the present work and it models the molecular dynamics of simple polymers. An important aspect of the Fokker-Planck dynamics in bistable systems is that the first excited state can become almost degenerate with the ground state especially for large activation barriers. The corresponding small, but non-zero, eigenvalue determines the rate of passage of particles from one potential well to the other, so defining a theoretical framework for a microscopic derivation of the rate coefficients of elementary two-state reactions as initiated by Kramers~\cite{Kramers1940, hanggi1990reaction}.
Although an accurate calculation of the first non-zero eigenvalue assumes a central relevance in the study of kinetic processes, it is computationally challenging especially for large systems characterized by many coupled coordinates. In this context, the application of new quantum computational tools can be convenient to possibly tackle previously inaccessible stochastic problems.\\

We propose to harness the exponential storage capacity of a quantum computer to handle this problem and we detail the implementation of the stochastic problem in the quantum architecture in Section~\ref{sec:implementation}. The eigenvalue problem associated to the Fokker-Planck operator is encoded into the quantum register with a simple binary mapping of the basis set in the computational basis. By exploiting the symmetry of the FPS operator, we discuss a VQE-like algorithm to obtain the first non-zero eigenvalue of the stochastic operator, corresponding to the kinetic rate of the isomerization reaction.\\

Section~\ref{sec:Results} presents the results obtained for the simulation of small chains characterized by different activation barriers. We discuss several aspects which play a critical role on the performance of the algorithm, such as the choice of the variational ansatz, the size of the basis set which is reflected on the number of qubits to be used and the effect of the noise on the accuracy of the result. In the concluding section, we critically discuss the potential quantum advantage of the proposed algorithm. The key advantage stems from the linear scaling in terms of memory resources, indeed the binary mapping implies a linear relation between the number of rotors in the polymeric chain and the number of qubits used in the implementation. This is an exponential gain compared to classical memory resources which grow with the dimension of the vector space representing the state of the system, dimension increasing exponentially with the number of relevant degrees of freedom. On the other hand, the present implementation suffers of an exponential scaling of the number of expectation values which needs to be measured to solve the problem and a somehow limited accuracy of the results for low barriers. We therefore point out future developments which are needed to bring the stochastic molecular dynamics amongst the applications which will take great advantage by the advent of quantum computers.

\section{Theory}\label{sec:Theory}
The stochastic description of molecular systems is a broad research field that embraces a wide range of theoretical tools to allow the interpretation of the fluctuating dynamics of microscopic systems in a plethora of physical conditions~\cite{van1992stochastic}. We will focus our attention on the simple case of a stationary and Markovian diffusive process. This kind of behavior is typical of the so called over-damped regime~\cite{gardiner1985handbook}, in which a time-scale separation is observed due to the configurational variables' dynamics being slower than the evolution of the conjugated momenta. By looking at the system within a time-scale comparable with the fluctuations of the configurational variables $\mathbf{q}(t)$, the momentum variables rapidly loose correlation, so that one can assume they are relaxed in the equilibrium distribution.\\

In order to describe the time evolution of the probability distribution $\rho(\mathbf{q},t)$, that depends on some configurational variables $\mathbf{q}$, the Fokker\hyp{}Planck\hyp{}Smoluchowski (FPS) equation can be invoked~\cite{gardiner1985handbook}:
\begin{equation}\label{FPS_equation}
    \frac{\partial \rho(\mathbf{q},t)}{\partial t} := - \hat{\Gamma}\rho(\mathbf{q},t)
\end{equation}
where the FPS operator $\hat{\Gamma}$ is defined as follows:
\begin{equation}\label{FPS_operator}
    \hat{\Gamma} = - \frac{\partial}{\partial \mathbf{q}}^T \mathbf{D}(\mathbf{q})\rho_{eq}(\mathbf{q})\frac{\partial}{\partial \mathbf{q}}\rho_{eq}(\mathbf{q})^{-1}
\end{equation}
Here $\mathbf{D}(\mathbf{q})$ is the diffusion tensor~\cite{gardiner1985handbook} and $\rho_{eq}(\mathbf{q})$ represents the Boltzmann equilibrium probability distribution:
\begin{equation}
    \rho_{eq}(\mathbf{q})\propto e^{-\beta U(\mathbf{q})}
\end{equation}
where $U(\textbf{q})$ is the mean-field potential and $\beta = (k_BT)^{-1}$.\\

The FPS operator is real and positive semi-definite such that each eigenvalue $\lambda_k$ obeys to the constraint $\lambda_k\geq0\;\forall k$. From now on we will consider the eigenvalues $\lambda_k$ sorted in ascending order such that $\lambda_k\leq\lambda_{k+1}$ and we will indicate with $\psi_k(\mathbf{q})$ the corresponding eigenfunctions.\\

The lowest eigenfunction $\psi_0(\mathbf{q})$, corresponding to the null eigenvalue $\lambda_0=0$, is provided by the equilibrium distribution $\rho_{eq}(\mathbf{q})$. If the time evolution of a generic non-equilibrium distribution $\rho(\mathbf{q},t)=\sum_{k} c_k(t)\psi_k(\mathbf{q})$ is considered, the following relation holds:
\begin{equation}\label{fps_dinamics}
    \rho(\mathbf{q},t+\tau)=e^{-\hat{\Gamma}\tau}\rho(\mathbf{q},t)
    =\sum_{k} e^{-\lambda_k \tau} c_k(t) \psi_k(\mathbf{q})
\end{equation}
From this equation we can easily conclude that, since all the non-zero eigenvalues play the role of exponential decay rates, a generic probability distribution must relax over time toward the equilibrium state $\psi_0(\mathbf{q})$. \\

The goal of this work is to investigate a possible strategy to solve, with the support of a quantum computer, the eigenvalue problem associated to the FPS operator from Eq.~\ref{FPS_operator}. This however cannot be done directly since the FPS operator is non Hermitian and, as such, cannot be converted into a qubit Hamiltonian to be used in the variational procedure. This issue has a simple work-around that involves the introduction of a symmetrized, self-adjoint operator $\hat{\Tilde{\Gamma}}$:

\begin{equation}\label{Simmetrized_FPS_operator}
    \hat{\Tilde{\Gamma}}:=\rho_{eq}(\mathbf{q})^{-\frac{1}{2}}\hat{\Gamma}\rho_{eq}(\mathbf{q})^{\frac{1}{2}}
\end{equation}
which leads to a "Schr{\"o}dinger-like" representation of the FPS operator \cite{elber2020molecular} with $\hat{\tilde{\Gamma}}$ in place of the Hamiltonian operator.\\

This new symmetrized operator conserves the same eigenvalues $\lambda_k$ of the original FPS operator while its eigenfunctions $\Tilde{\psi}_k(\mathbf{q})$ are connected with the eigenfunctions $\psi_k(\mathbf{q})$ of $\hat{\Gamma}$ by the relation:
\begin{equation}
    \Tilde{\psi}_k(\mathbf{q}) = \rho_{eq}(\mathbf{q})^{-\frac{1}{2}} \psi_k(\mathbf{q})
\end{equation}
For more details about the symmetrized FPS operator we refer to~\cite{gardiner1985handbook, jungel2016fokker}.\\

Here, inspired by the Variational Quantum Eigensolver (VQE)~\cite{Peruzzo2014}, we propose an hybrid algorithm to obtain the first non-vanishing eigenvalue (\textit{i.e.}, $\lambda_1$) of $\hat{\Tilde{\Gamma}}$. The procedure exploits the quantum computer to prepare, by means of a parametrized circuit, a trial wave-function to approximate the state correspondent to the $\tilde{\psi}_1(\mathbf{q})$ target eigenfunction. Subsequently an external classical optimization routine updates the parameters entering the variational circuit on the basis of the measured expectation value $\langle \hat{\Tilde{\Gamma}} \rangle$. The procedure is iterated until convergence. In order to measure $\langle \hat{\Tilde{\Gamma}} \rangle$ the symmetrized FPS operator will be represented as a linear combination of Pauli strings.

\begin{equation} \label{pauli_sum}
\begin{gathered}
    \hat{\Gamma}_{QC} = \sum_{j} \gamma_j \hat{P}_j \\
    \langle \hat{\Tilde{\Gamma}} \rangle \equiv \langle \hat{\Gamma}_{QC} \rangle = \sum_{j} \gamma_j \langle \hat{P}_j \rangle 
\end{gathered}
\end{equation}
where the explicit formulation for the weights $\gamma_j$ and the Pauli strings $\hat{P}_j \in \{ \sigma_x, \sigma_y, \sigma_z, \mathbb{I} \}^{\otimes N}$ depends upon the adopted mapping (see Sec.~\ref{sec:implementation}).\\

Notably, the direct application of the VQE to the FPS eigenvalue problem would lead to the trivial solution $\lambda_0=0$, as already mentioned. In the following (Sec. \ref{subsec:Model}), we will report the implementation of a particular problem for which the symmetry allows to easily target the desired eigenvalue. Nevertheless, various techniques can be applied to overcome this issue in analogy with  the computation of excited states of a molecular Hamiltonian~\cite{colless2018computation, ollitrault2020quantum, higgott2019variational}. 

In the following section we will present, on the basis of this theoretical introduction, a linear chain molecule as model system to implement this procedure. We will focus on the calculation of the first excited state that, as will be discussed, assumes a central relevance in the study of kinetic processes.

\subsection{Stochastic dynamics and conformational transition rate}\label{subsec:Model}

The FPS equation describes the time evolution of the probability distribution over all the configurational domain allowing for a continuous description of the system dynamics. This degree of detail is surely very informative, but is also not always the best way of describing a reactive process. Often, either for simplicity or due to the lack of information about the initial distribution, less detailed descriptions are adopted, in which, rather than describing the molecular configurations, one usually thinks to the process as the inter-conversion between reactant and product structures~\cite{elber2020molecular}. In doing so we implicitly moved our description from a continuous FPS problem to a discrete Master equation with rates identifying the kinetic constants for the interconversions between different states.\\

\begin{figure}[h!]
    \centering
    \includegraphics[width = \columnwidth]{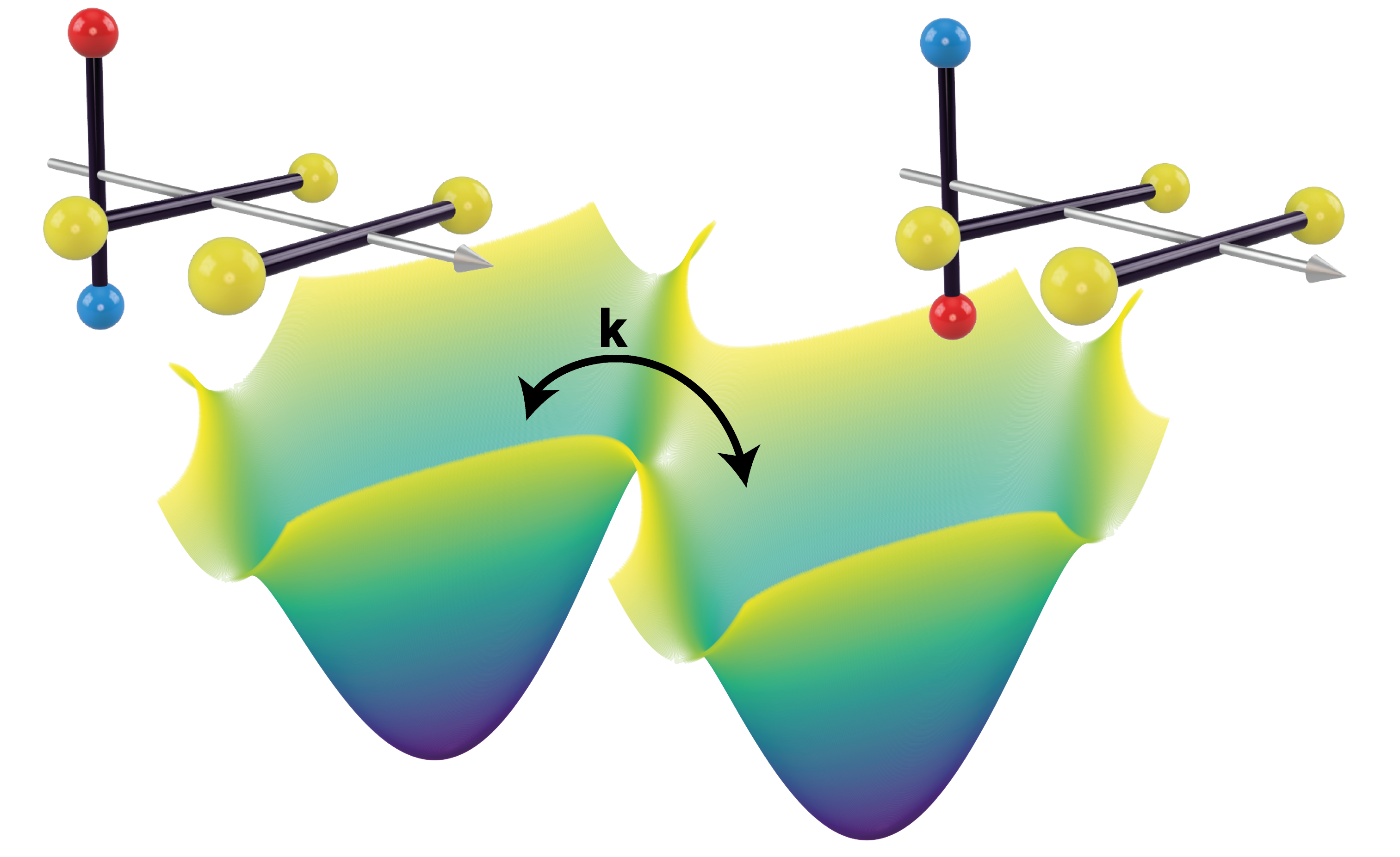}
    \caption{Pictorial representation of the isomerization process in the chain of three rotors. The potential energy surface corresponds to the two different configurations of the chain molecule, the kinetic rate constant for the process is determined by the first non-zero eigenvalue of the Fokker-Planck-Smoluchowski operator. }
    \label{buche_bacchette}
\end{figure}

In this paper, we will consider a symmetrical double minimum system in which two stable molecular configurations are kept apart by a potential barrier. This is the prototypical example of a simple isomerization process in which the kinetic constants $k$ of the direct and reverse processes are the same. If the Master equation description is adopted, it is trivial to show how the population relaxation rate toward equilibrium is described by the exponential decay $\exp{(-2kt)}$. If the barrier is large enough, the FPS operator eigenvalue spectrum shows a large gap between the first non-zero eigenvalue $\lambda_1$ and the higher ones. In this picture, the states corresponding to $\lambda_n\gg \lambda_1$ have the character of fast (intra-minimum) relaxing modes, while the longer-lived state, corresponding to $\lambda_1$, represents the kinetic relaxation mode connected with the population transfer between sites (Figure~\ref{buche_bacchette}). The exponential form dictating the relaxation dynamic can be obtained from Eq.~\ref{fps_dinamics} as $\exp{(-\lambda_1t)}$ from which the simple relation, $\lambda_1 = 2k$, can be established between the kinetic constant $k$ and the first non-vanishing FPS operator eigenvalue $\lambda_1$.\\

\subsection{The rotor chain as a model for molecular conformational dynamics}

\begin{figure}[b!]
\centering
\tdplotsetmaincoords{75}{130}
\begin{tikzpicture}[scale=2, tdplot_main_coords]
    \draw[-latex, thick] (0, 0, 0) -- (1, 0, 0);
    \draw[-latex, thick] (0, 0, 0) -- (0, 3, 0);
    \draw[-latex, thick] (0, 0, 0) -- (0, 0, 1);
    \draw[dashed, gray] (0, 0.5, 0) -- (0, 0.5, 1);
    \draw[dashed, gray] (0, 1.5, 0) -- (0, 1.5, 1);
    \draw[dashed, gray] (0, 2.5, 0) -- (0, 2.5, 1);
    \draw[dotted, red] (0.423, 0, 0.906) -- (0.423, 2, 0.906);
    \draw[dotted, blue] (-0.5, 1, 0.866) -- (-0.5, 3, 0.866);
    \draw (0, 0.5, 0) node[circle, fill, inner sep=1pt,label=below:$1$](){};
    \draw (0, 1.5, 0) node[circle, fill, inner sep=1pt,label=below:$2$](){};
    \draw (0, 2.5, 0) node[circle, fill, inner sep=1pt,label=below:$3$](){};
    \draw[-latex, red, thick] (0, 0.5, 0) -- (0.423, 0.5, 0.906);
    \draw[-latex, blue, thick] (0, 1.5, 0) -- (-0.5, 1.5, 0.866);
    \draw[-latex, green!50!black, thick] (0, 2.5, 0) -- (0.785, 2.5, 0.785);
    \draw[red, dashed] (0, 1.5, 0) -- (0.423, 1.5, 0.906);
    \draw[blue, dashed] (0, 2.5, 0) -- (-0.5, 2.5, 0.866);
    \tdplotsetthetaplanecoords{0}
    \tdplotdrawarc[tdplot_rotated_coords, red]{(0,0,0.5)}{0.4}{0}{25}{anchor=south}{$\varphi_0$}
    \tdplotdrawarc[tdplot_rotated_coords, blue]{(0,0,1.5)}{0.4}{0}{-30}{anchor=south}{$\varphi_1$}
    \tdplotdrawarc[tdplot_rotated_coords, green!50!black]{(0,0,2.5)}{0.4}{0}{45}{anchor=south}{$\varphi_2$}
    \tdplotdrawarc[tdplot_rotated_coords,latex-latex]{(0,0,1.5)}{0.8}{25}{-30}{anchor=south}{$\theta_1$}
    \tdplotdrawarc[tdplot_rotated_coords,latex-latex]{(0,0,2.5)}{0.8}{45}{-30}{anchor=south}{$\theta_2$}

\end{tikzpicture}
\caption{Schematic representation of a chain of three rotors reporting the definition of the set of $\varphi_k$ orientations and of the corresponding set of dihedral angles $\theta_k$.} \label{fig:RotorChainTIKZ}
\end{figure}
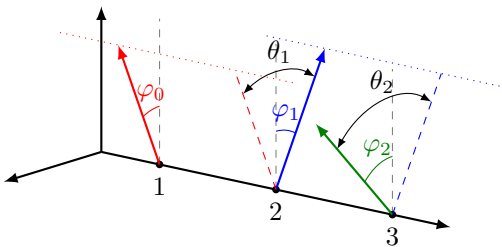

Now that the theoretical framework has been outlined let us introduce a model system to be used as a test subject. Let us consider a chain of $N+1$ rotors free to rotate around a common axis, see Fig.\ref{fig:RotorChainTIKZ}. Let $\varphi_k$ (for $k = 0 \, ,  1 \, , \dots, N$) be the angular coordinate representing the orientation of the $k$-th rotor with respect to a laboratory axis orthgonal to the chain axis and $\theta_k:=\varphi_{k}-\varphi_{k-1}$ the relative orientation angles, hereafter called as dihedral angles, between the $k$-th rotor and the previous one in the chain.

Let us assume that the rotors interact on the basis of a nearest-neighbour periodic potential $U_k(\theta_k) = U_k(\theta_k + 2\pi)$, depending on the corresponding dihedral angle $\theta_k$,  such that the overall mean-field potential for the chain can be specified as:
\begin{equation}\label{Total_Potential}
    U(\bm{\theta}):=\sum_{k=1}^N U_k(\theta_k)
\end{equation}
in which $\bm{\theta}=\left( \theta_1, ..., \theta_N \right) \in \mathbb{R}^N$ represents the vector encoding the internal configuration of the chain. 

In order to have a clear picture of the kinetic process, let us consider a chain in which $N-1$ mono-stable dihedrals are coupled to a single bi-stable potential. In what follows the mono-stable potential will be specified as:
\begin{equation}\label{mono-stable potential}
    U_k(\theta_k):=\frac{\Delta_\text{nr}}{2}\left[1-\cos{(\theta_k)}\right]
\end{equation}
while the bi-stable one by the form:
\begin{equation}\label{bi-stable potential}
    U_k(\theta_k):=\frac{\Delta_\text{r}}{2}\left[\cos{(2\theta_k)}+1\right]
\end{equation}
where we have adopted the symbols $\Delta_\text{r}$ and $\Delta_\text{nr}$ to indicate the barrier height encountered when moving, respectively, along the \textit{reactive} and \textit{non-reactive} dihedral coordinates.\\

Furthermore, let us assume the absence of hydrodynamic interactions~\cite{doi1988theory} amongst the rotors which, therefore, are considered as characterized by independent and constant diffusion coefficients ($D_k$ for $k = 0 \, ,  1 \, , \dots, N$). Under these assumptions the FPS operator for the process can be written as:
\begin{equation}\label{FPS_op_chain_rotors}
    \begin{split}
        \hat{\Tilde{\Gamma}}&=-\sum_{k=0}^{N}D_k\rho_{eq}^{-\frac{1}{2}}(\bm{\varphi})\frac{\partial}{\partial \varphi_k}\rho_{eq}(\bm{\varphi})\frac{\partial}{\partial \varphi_k}\rho_{eq}^{-\frac{1}{2}}(\bm{\varphi})\\
    \end{split}
\end{equation}
in which $\bm{\varphi}=\left(\varphi_0,..., \varphi_{N}\right)\in\mathbb{R}^{N+1}$ represents the vector encoding the orientation of each rotor.

The FPS operator defined in Eq.~\ref{FPS_op_chain_rotors} contains a degenerate degree of freedom. Because of the pairwise decomposition of the mean field potential (see Eq. \ref{Total_Potential}), any homogeneous rotaion of the rotors brings the system into an equivalent conformational state. Such a degeneracy can be exploited by introducing a coordinate representation based on the dihedral angles $\bm{\theta}$ and the overall chain orientation $\Phi$ defined as:

\begin{equation}
    \Phi := \frac{\sum_{k=0}^N \frac{\varphi_k}{D_k}}{\sum_{k=0}^N \frac{1}{D_k}} 
\end{equation}

If the coordinate transformation $\bm{\varphi}\rightarrow \left[\Phi, \bm{\theta}\right]$ is applied, the FPS operator can be decomposed as the sum of a global orientation operator $\hat{\tilde{\Gamma}}_\Phi$, $N$ single dihedral terms $\hat{\tilde{\Gamma}}_k$ and $N-1$ two dihedrals interaction terms $\hat{\tilde{\Gamma}}_{k, k+1}$:
\begin{equation}
    \hat{\tilde{\Gamma}}\rightarrow\hat{\tilde{\Gamma}}_\Phi+\sum_{k=1}^N\hat{\tilde{\Gamma}}_k+\sum_{k=1}^{N-1}\hat{\tilde{\Gamma}}_{k, k+1}
\end{equation}
with these operators specified as:
\begin{align}
    &\hat{\tilde{\Gamma}}_\Phi = -\frac{1}{\sum_{k=0}^N D_k^{-1}}\frac{\partial^2}{\partial\Phi^2}\\
    &\hat{\tilde{\Gamma}}_k = -\left(D_k+D_{k-1}\right)\left[\frac{\partial^2}{\partial\theta_k^2}+\frac{U_k''(\theta_k)}{2}-\frac{U_k'(\theta_k)^2}{4}\right]\\
    &\hat{\tilde{\Gamma}}_{k,k+1} =2D_k\left[ \frac{\partial}{\partial\theta_k}\frac{\partial}{\partial\theta_{k+1}}-\frac{1}{4}U_k'(\theta_k)U_{k+1}'(\theta_{k+1})\right]
\end{align}
In order to study the conformational dynamics, the term $\hat{\tilde{\Gamma}}_\Phi$, depending on the global orientation angle $\Phi$, can be neglected since it does not affect the internal degrees of freedom.\\

At this point, in order to solve the eigenvalue problem associated with the FPS operator and to obtain the integrals required for the VQE algorithm, a convenient basis set must be selected. For the purpose of this paper a set $\{\phi_\mathbf{n}(\bm{\theta})\}$ of composite basis functions has been constructed according to:
\begin{equation}\label{composite_basis_set}
    \phi_\mathbf{n}(\bm{\theta})=\prod_{k=1}^N \Xi_{n_k}(\theta_k)    
\end{equation}
where $\Xi_{n_k}(\theta_k)$ are the eigenfunctions of the single dihedral operator $\hat{\tilde{\Gamma}}_k$ and $\mathbf{n}=(n_1,...,n_N)$ represents a vector containing the order of the eigenfunctions for each dihedral angle. All the isolated dihedral eigenfunctions $\Xi_{n_k}(\theta_k)$ have been obtained by expanding the proper FPS operator over a Fourier basis set. 

Given the parameterized forms of Eq. \ref{mono-stable potential} and Eq. \ref{bi-stable potential} for the contributions to the mean field potential, one derives quite easily that the FPS operator $\hat{\tilde{\Gamma}}$ is invariant with respect to the change of sign of all the dihedral angles.
In mathematical terms this condition implies the commutation relation $\left[\hat{\Gamma},\hat{S}\right]=0 $, where $\hat{S}$ is the symmetry operator for the change of sign of the dihedral angles. In essence the FPS operator acts on a symmetry parted Hilbert space $\mathcal{H} = \mathcal{H}_{+} \oplus \mathcal{H}_{-}$ and its action does not mix functions belonging to the even $\mathcal{H}_{+}$ and to the odd $\mathcal{H}_{-}$ sub-spaces. So, the variational theorem applies to the lowest eigenvalue of both sub-spaces. Therefore both the first eigenstate $\Tilde{\psi}_0(\mathbf{q}) = \rho_{eq}(\mathbf{q})^{-\frac{1}{2}}$ and the second eigenstate $\tilde{\psi}_1$, associated with the aforementioned kinetic mode, can be obtained from the variational procedure applied to the even and odd sub-spaces, respectively. In this framework, the application of a VQE procedure to get $\lambda_1$ requires the representation of FPS operator on the basis of odd basis functions only.  

Given $p_{n_k}\in(-1,1)$ as the parity index of each dihedral eigenfuntion $\Xi_{n_k}(\theta_k)$, the condition:
\begin{equation}\label{composite_basis_set_rule}
    \prod_{k=1}^N p_{n_k} = -1 \qquad \forall \mathbf{n} \mid \phi_\mathbf{n}(\bm{\theta})\in\mathcal{H}_-
\end{equation}
must be verified in order to consider only functions belonging to the odd $\mathcal{H}_-$ sub-space. The results from the classical computations of the stochastic problem will be discussed, together with the ones coming from the VQE procedure, in Sec.~\ref{sec:Results}.

\section{Implementation}\label{sec:implementation}
The first step in the implementation is represented by the translation of the target $\Tilde{\psi}_1(\mathbf{q})$ eigenfunction from a linear combination of $N$-dihedral states (Eq.~\ref{composite_basis_set}) to a linear combination of quantum computer bitstring states. Such a procedure will provide also an expression for the symmetrized FPS operator as a combination of Pauli strings. In this paper, we opted for a binary mapping in which a progressive numerical label $j$ is assigned to every basis function $\phi_\mathbf{n}(\bm{\theta})$. Each label is then directly mapped through its binary representation $\text{bin}(j)$ in a state $\lvert j \rangle\equiv\lvert \text{bin}(j) \rangle$ of the qubit register. Given a set of $J$ basis functions a total of $Q=\lceil \log_2(J) \rceil$ qubits is required.\\

We start the mapping process by considering the FPS operator representation in the previously introduced symmetry adapted basis:
\begin{equation}\label{FPS_operator_projectors}
    \hat{\tilde{\Gamma}}=\sum_{r,c} \gamma_{r,c}\lvert r \rangle \langle c \rvert \quad \text{with} \quad \gamma_{r,c}=\langle r \rvert \hat{\tilde{\Gamma}} \lvert c \rangle
\end{equation}
Let us indicate with $\delta_q(j)$ the $q$-th digit of the binary representation of $j$ such that $\lvert j \rangle\equiv\lvert\delta_1(j),...,\delta_Q(j)\rangle$, we can rephrase the operator representation, Eq.~\ref{FPS_operator_projectors}, in a qubit-wise form:
\begin{equation}
    \hat{\tilde{\Gamma}}=\sum_{r,c} \gamma_{r,c} \bigotimes_{q=1}^Q\lvert \delta_q(r) \rangle \langle \delta_q(c) \rvert
\end{equation}
By substituing in the last equation each single qubit outer product as combination of Pauli matrices, we obtain the following decomposition in a linear combination of Pauli strings like in Eq. \ref{pauli_sum}:
\begin{equation}\label{Binary-mapping}
    \begin{split}
        \hat{\tilde{\Gamma}}&=\sum_{r,c}\gamma_{r,c} \bigotimes_{q=1}^Q \left\{\frac{1-\left|\delta_q(r)-\delta_q(c)\right|}{2}\left[\bm{1}_q + (-1)^{\delta_q(r)}\bm{\sigma}_q^z\right] \right. + \\
        &\left. \qquad+ \frac{\left| \delta_q(r)- \delta_q(c) \right|}{2}\left[ \bm{\sigma}_q^x + (-1)^{\delta_q(r)} i\bm{\sigma}_q^y \right]\right\}
    \end{split}
\end{equation}
As we can see each matrix element of the FPS operator is translated into a set of $2^Q$ strings, each of which is composed by $Q$ Pauli matrices. Each set of strings is specific for the given binary bitstring representation of all the possible $r$ and $c$ states.\\

Here we discuss the potential quantum advantage of the approach. Considering that the number of basis states $J$ to encode shows a roughly exponential increase with the system size (e.g. with the length of the rotor chain), the adoption of this mapping protocol allows to overcome the exponential growth of a classical storage space by requiring a number of qubits which scales linearly with the system size. To be more precise, let us indicate with $m$ a positive integer related to the number of single-dihedral basis states used to set up the product states in Eq. \ref{composite_basis_set} and let $N$ be the number of dihedral angles in the chain. From these assumptions we can estimate the order of $J$ to be $J\sim\mathcal{O}(m^N)$ and, consequently, only $\mathcal{O}(N \log_2(m))$ qubits are needed to represent the system on the quantum register.
In order to fully characterize the efficiency associated with the procedure, a more in depth analysis of the mapping is required. To do so, the measurement process has to be taken into account.  As mentioned in Sec. \ref{sec:Theory}, the expectation value measurement is related, according to Eq.~\ref{pauli_sum}, to the expectation values of the Pauli strings which, in turn, are affected both by the structure of the operator and the mapping adopted. Since we have already discussed the features of the mapping now we should turn our attention to the estimation of the non-vanishing elements of the FPS operator.\\

Due to the nearest-neighbour nature of the interaction the $\hat{\tilde{\Gamma}}$ operator is sparse. Indeed, it has a block structure where only states differing for single-dihedral function of adjacent rotors are connected. Nevertheless, even with this great reduction, the number of Pauli strings to be measured is exponentially growing: the number of elements connecting states that differ for two single-dihedral functions is $\mathcal{O}(m^2JN)$ to which we add the $\mathcal{O}(mJN)$ elements differing for a single-dihedral state and the $J$ diagonal elements. It may be worth to notice that this estimate does not take into account any symmetry argument that could reduce the count and does not consider additional strategies that can be employed to reduce even more the number of required measurements. Some consideration about this point will be discussed in Sec.~\ref{sec:Conclusions}.\\

It is important to notice that this is not the only possible choice for the mapping and different strategies could be more or less convenient depending on the system at hand; ref.~\cite{sawaya2020resource} details several possibilities available within the range set by the binary mapping (more dense) and the unary encoding (less dense, where $d$-level systems are mapped into $d$ bitstring states with all the qubits in the state $|0\rangle$ but one distinguishing among different states, see for instance refs.~\cite{mcardle2019digital, castaldo2021quantum}). As a general rule, a more or less dense mapping results in an accordingly higher or smaller cost in terms of length and number of Pauli strings generated by each mapped matrix element. These different factors must be balanced according to the system and its size. 

That said, the binary mapping suffices for our scope which is to give a proof of concept of the application by considering a rather small system. The design of a more efficient mapping is left for future development and discussed in Sec. \ref{sec:Conclusions}.

\subsection{Computational details}\label{subsec:Computational_details}

Here we provide the details about the numerical results that are shown in Sec.~\ref{sec:Results}. The matrix elements of the FPS operator have been obtained with a homemade C\texttt{++} code~\cite{smoluchowski_rotor_chain} which has also been used as benchmark for the implementation of the hybrid algorithm in a Python code~\cite{binary_vqe}. 

In order to evaluate the accuracy and the performance of our method we have considered a chain of three rotors. The two resulting dihedral angles have been parametrized as follows: the first one experiences a bi-stable potential, as dictated by Eq.~\ref{bi-stable potential}, while the second one is described by a mono-stable potential as from Eq.~\ref{mono-stable potential}. The barrier height $\Delta_\text{r}$ for the bistable dihedral has been varied from $0.5\,k_BT$ to $3.0\,k_BT$ while the $\Delta_\text{nr}$ parameter for the single-minimum dihedral has been kept fixed at $1\,k_BT$. The diffusion coefficients for all three rotors have been set to the same relative value of $1$. The term "relative" is here employed since, due to the structure of Eq.~\ref{FPS_op_chain_rotors}, any scale factor can be applied to the FPS operator to scale the result in term of the desired diffusion coefficient.

The algorithm has been tested on $Q = 2, 3, 4$ qubits quantum registers using a composite basis set filling completely all the $2^Q$ register states. In order to generate the proper composite basis set for the system an adequate number of single dihedral basis functions had to be selected according to Eq.~\ref{composite_basis_set} and Eq.~\ref{composite_basis_set_rule}. Indicating the number $n_1$ of isolated basis function for the double minimum dihedral and with $n_2$ the one selected for the single minimum one, the following $(n_1,n_2)$ basis set have been selected for each value of $Q$: For the $2$ qubit system a $(4,2)$ basis set has been employed, for the $3$ qubit we opted for a $(4, 4)$ configuration and a $(8, 4)$ set has been used for the $4$ qubit calculations. Note that, due to symmetry restrains, only half of the composite basis functions generated by each $(n_1,n_2)$ set are used in the actual variational procedure.\\

\begin{figure}[b!]
    \begin{subfigure}{\columnwidth}
        \centering
        \includegraphics[width=\columnwidth]{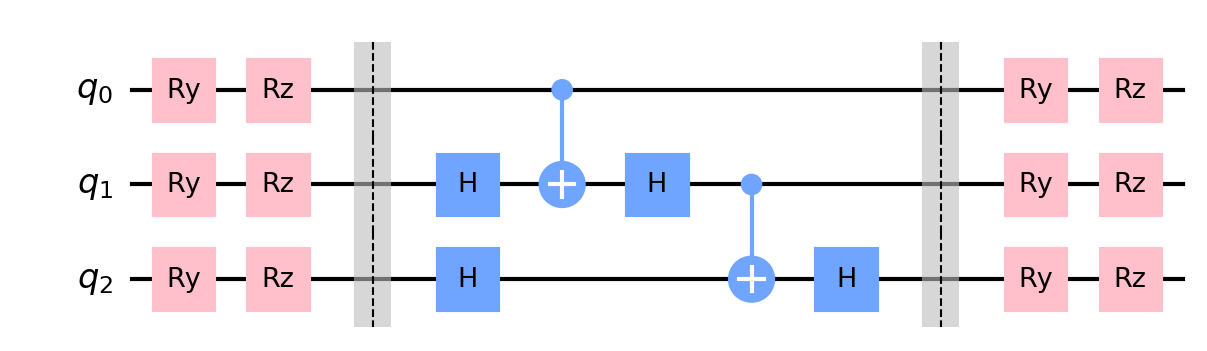}
        \caption{Linear entangler}
        \label{subfig:linear_RyRz}
    \end{subfigure}
    \begin{subfigure}{\columnwidth}
        \centering
        \includegraphics[width=\columnwidth]{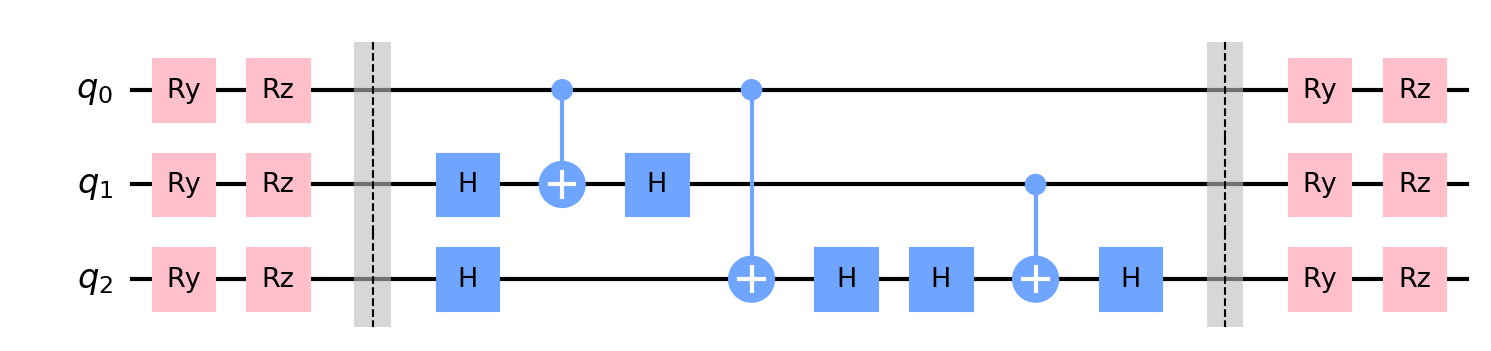}  
        \caption{Full entangler}
        \label{subfig:full_RyRz}
    \end{subfigure}
    \caption{RyRz variational forms of depth $d=1$ with linear (a) or full (b) entangler blocks on a 3-qubits circuit.}
    \label{fig:RyRz}
\end{figure}

A RyRz heuristic variational circuit has been employed in order to prepare the trial wave-function required in the VQE procedure~\cite{moll2018quantum}. Two main elements form such a circuit: a layer of parameterized $R_y$ and $R_z$ rotations, used to apply an arbitrary single qubit rotation to each qubit in the quantum register, and an entangler block to introduce correlation in the trial wavefunction. A variational circuit of depth $d=1$ can be obtained enclosing an entangler block between two rotation layers. Higher depth circuits can be obtained by concatenating couples of entangler blocks and rotation layers to the depth $d=1$ circuit. A $Q$-qubits RyRz circuit of depth $d$ is defined by $2Q(d+1)$ variational parameters. Two different entangler blocks configurations have been considered in this paper: one creating entanglement between adjacent qubits, hereafter addressed with the name "linear", and one creating entanglement between all couples of qubits in the register, called "full" from now on. In Figure~\ref{fig:RyRz} a schematic representation of these parameterized circuits is shown.\\

In order to start the VQE procedure the register is set to the all zeros state $\lvert 0 \rangle$ that has been selected in order to correspond to the basis set function $\phi_0(\theta_1,\theta_2)=\Xi_1(\theta_1)\Xi_0(\theta_2)$ given by the product of the first excited state $\Xi_1(\theta_1)$ for the isolated double minimum system and the ground-state $\Xi_0(\theta_2)$ of the single minimum one. The rotation angles of the variational form are set randomly at every run.\\ 

Four different optimization algorithm, namely the downhill simplex method, Sequential least square programming (SLSQP), Constrained optimization by linear approximation (COBYLA) and Simultaneous perturbation stochastic approximation (SPSA), have been applied with a variable degree of success based on the type of simulation. The IBM's Qiskit module~\cite{Aleksandrowicz2019} (version 0.22) has been employed for the simulation of the quantum circuits. The implementation has been verified using the built-in Qiskit Statevector Simulator, an ideal simulator that return the statevector of the quantum register. Then, we characterized the statistics induced by a finite number of measurements of the quantum state in a noiseless quantum simulation varying the number of runs (that are the times a circuit is executed and measured). Finally the behavior of the algorithm has been tested on a noisy simulator. For this last task, the noise model of the IBMQ Santiago has been employed and a basic readout error mitigation protocol (based on Ref.~\cite{fiuravsek2001maximum}) has been applied to each circuit simulation as implemented in Qiskit. \\

In order to perform the simulations with the IBMQ Santiago noise model, the built-in Qiskit transpilation routines have been employed to map the circuit on the quantum computer topology. During the process the gates have been translated into $U_1$, $U_2$, $U_3$ and CNOT basis gates. Given the 5 inline qubits topology of the IBMQ Santiago quantum computer the transpilation of a linear entangler RyRz variational form on $Q$ qubits appears to be very efficient resulting in a gate count of $4Q-3$ ($3Q-2$ $U_n$ and $Q-1$ CNOT) without requiring any qubit swap. The transpilation of a full entangler RyRz circuit represent instead a more complex task requiring qubit swaps due to the presence of CNOT entagling operations between non-adjacent physical qubits.

The expectation value measurement has been carried out either by measuring independently all the Pauli strings of the mapped operator or by retrieving from Qiskit the built-in method which groups all the terms that are simultaneously diagonalizable~\cite{bravyi2017tapering}. Both methods performed equally well and returned comparable results.

\section{Results}\label{sec:Results}

In this section we discuss the results obtained with numerical simulations of the algorithm in several platforms: the Qiskit Statevector Simulator, the Qiskit Quantum Assembly Simulator (QASM) with and without a noise model and the IBMQ Santiago device~\cite{Aleksandrowicz2019}.

In Sec.~\ref{subsec:variational_network} we assess the variational circuit performance by discussing the results obtained with a noise-free optimization. The importance of this question is highlighted by the large amount of research on the ansatze efficiency when this kind of algorithms are applied to quantum chemistry problems~\cite{wecker2015progress, babbush2018low, Peruzzo2014}. Finally, in Sec. \ref{subsec:Noisy_results} we show the results of noisy simulations of the algorithm as a function of the barrier height and number of qubits adopted.


\subsection{Variational network assessment}\label{subsec:variational_network}

The first question we want to address is whether a RyRz variational form can produce a good trial wavefunction to approximate the classical computational result and if the optimization procedure is able to find its way towards such a configuration.\\ 

In order to give a reasonably general picture a few representative results have been reported in Tab.~\ref{tab:table_1}. All the data in Tab.~\ref{tab:table_1} have been obtained from $60$ independent VQE optimizations carried out on the Qiskit statevector simulator. For each evaluation the parameters of the RyRz variational form have been set randomly and optimized with the SPSA algorithm for a maximum of $600$ iterations. The choice of the optimization algorithm has been dictated by a performance analysis carried out on different optimizers reported in appendix A. The expectation value has been evaluated at each step with an independent measurement for each of the Pauli strings. \\

We have chosen to report the minimum value obtained from the sampling as a measure of the heuristic network capability in generating a trial state reasonably close to the optimal configuration; while we have adopted the average converged result to quantify the performance of the optimizer in extracting the desired state from the overall accessible Hilbert space. We are aware that this is a non trivial analysis given that many factors contribute to the shape of the optimization landscape and the exploration of the underlying Hilbert space. Nevertheless, we will see as our considerations are in agreement with more structured studies in literature applied to parametrized quantum circuits~\cite{woitzik2020entanglement, sim2019expressibility}.\\


\begin{table*}[t]
    \small
    \centering
    \caption{Results obtained by 60 independent VQE optimizations using the statevector simulations. $Q$ indicates the number of qubits and $\Delta_\text{r}$ is the barrier height for the double mimimum dihedral. We report $\min(\lambda_1)$ as the minimum value obtained in the simulations, $\langle \lambda_1 \rangle$ as the average value, while $\lambda_1^\text{ref}$ represents the theoretical target value for the considered basis set obtained with classical calculation. $\varepsilon_{min}(\%)$ and $\varepsilon_{avg}(\%)$ are respectively the percentage error on the minimum and on the average.}
    \label{tab:table_1}
    \begin{tabular*}{\textwidth}{@{\extracolsep{\fill}}ccc|ccc|cc}
        \hline
        $Q$ & $\Delta_\text{r}$ & Entangler & $\min(\lambda_1)$ & $\langle\lambda_1\rangle$ & $\lambda_1^\text{ref}$ & $\varepsilon_{min}(\%)$ & $\varepsilon_{avg}(\%)$ \\
        \hline
        2 & 0.5 & Linear & 1.51615 & 1.55182 & 1.51562 & 0.0350\% & 2.39\% \\
        3 & 0.5 & Linear & 1.48244 & 1.56413 & 1.47537 & 0.479\% & 6.02 \% \\
        3 & 0.5 & Full & 1.49113 & 1.56967 & 1.47537 & 1.07\% & 6.39\% \\
        4 & 0.5 & Linear & 1.54594 & 1.81958 & 1.47531 & 4.79\% & 23.34\% \\
        2 & 3.0 & Linear & 0.33313 & 0.33994 & 0.33310 & 0.00901\% & 2.05\% \\
        \hline
    \end{tabular*}
\end{table*}

As it can be seen from Tab.~\ref{tab:table_1} increasing the number of qubits results in a steep rise of the average value error accompanied by a somewhat milder increase in the error on the minimum value that, even for $4$ qubits, remains under the $5\%$ limit. These data indicate how a RyRz variational form of depth $d=1$ can be used to produce reasonable approximations of the target state with an error, on the expectation value, lower than $1.1\%$ for $2$ and $3$ qubits systems. This finding is in agreement with the general analysis carried out by Sim. \textit{et al.}~\cite{sim2019expressibility} where parametrized quantum circuits similar to the ones adopted in this study have shown a good capability of uniformly represent the Hilbert space in which they are defined.

Conversely, the higher error on the distribution average is a clear sign of the challenging optimization process. It is known that such a complexity is not only arising from the dimensionality of the problem but it is strictly related to the phenomenon of quantum tipicality~\cite{popescu2006entanglement, fresch2013typical} meaning that the distribution of certain functions over the quantum states of a given Hilbert space is extremely peaked around a typical value. Specifically, for random quantum circuits such as the one considered in our case, the average value of the gradient objective function tends to zero and as the Hilbert dimension increases the more states will correspond to flat optimization landscape regions~\cite{mcclean2018barren, cerezo2021cost}.

In order to clarify how the optimizer convergence impacts the overall VQE outcome let us observe how a state $\lvert \phi \rangle$ obtained from a $N$-qubit RyRz variational form can be exactly reproduced, in the form of $\lvert \phi \rangle \otimes \lvert 0 \rangle$, by a $(N+1)$-qubit RyRz variational ansatz by properly setting the rotation parameters. This implies that: given two variational systems, composed respectively by $N$ and $N'>N$ qubits, and considering the proper ordering of the basis-set, the accuracy of the best estimate accessible to the larger $N'$-qubits system must be higher or at least equivalent to the accuracy of the estimate produced by the smaller $N$-qubits one. In order to test this idea a hierarchical procedure has been devised in which the optimized parameters of a $N$-qubit variational form have been used as the starting guess in a $(N+1)$-qubit VQE procedure. A total of $500$ VQE have been carried out using COBYLA as the optimizer. Differently from what observed in a regular VQE procedure, as the one reported in Tab.~\ref{tab:table_1}, a monotonic decrease in the minimum value is encountered in the hierarchical procedure; a minimum value of $1.516$ is found for a $2$ qubits system, a value of $1.484$ is encountered for a $3$ qubits system while a final value of $1.475$ is encountered in the case of $4$ qubits. A similar monotonic decrease is also observed in the case of the distribution average. We therefore verified that the accuracy of the obtained result increases by enlarging the computational space in the case the optimization procedure is guided by an educated guess.

This observation is fundamental in understanding the nature of the results discussed above proving that the algorithmic bottleneck for the accuracy is the classical optimization procedure rather than the quantum portion. In conclusion, the trends emerging from these calculations are supporting the following considerations: the pure state distribution generated by the two entanglers allows one to reach the configuration corresponding to the global minimum of the optimization landscape but, on the other hand, we have seen that the landscape itself is affected by the dimension of the explorable Hilbert space. This drawback could be overcome developing physically-inspired ansatze for the Smoluchowski operator similarly to what has been done in literature for the molecular Hamiltonian~\cite{wecker2015progress}.\\


The results presented in this section will be useful in the following discussion: knowing the limit imposed by an ideal implementation will allow us to independently verify how various types of noise affect the overall stability of the method.


\subsection{Statistics of finite measurements and quantum noise effects}\label{subsec:Noisy_results}

\begin{figure}[h!]
    \centering
    \includegraphics[width=\columnwidth]{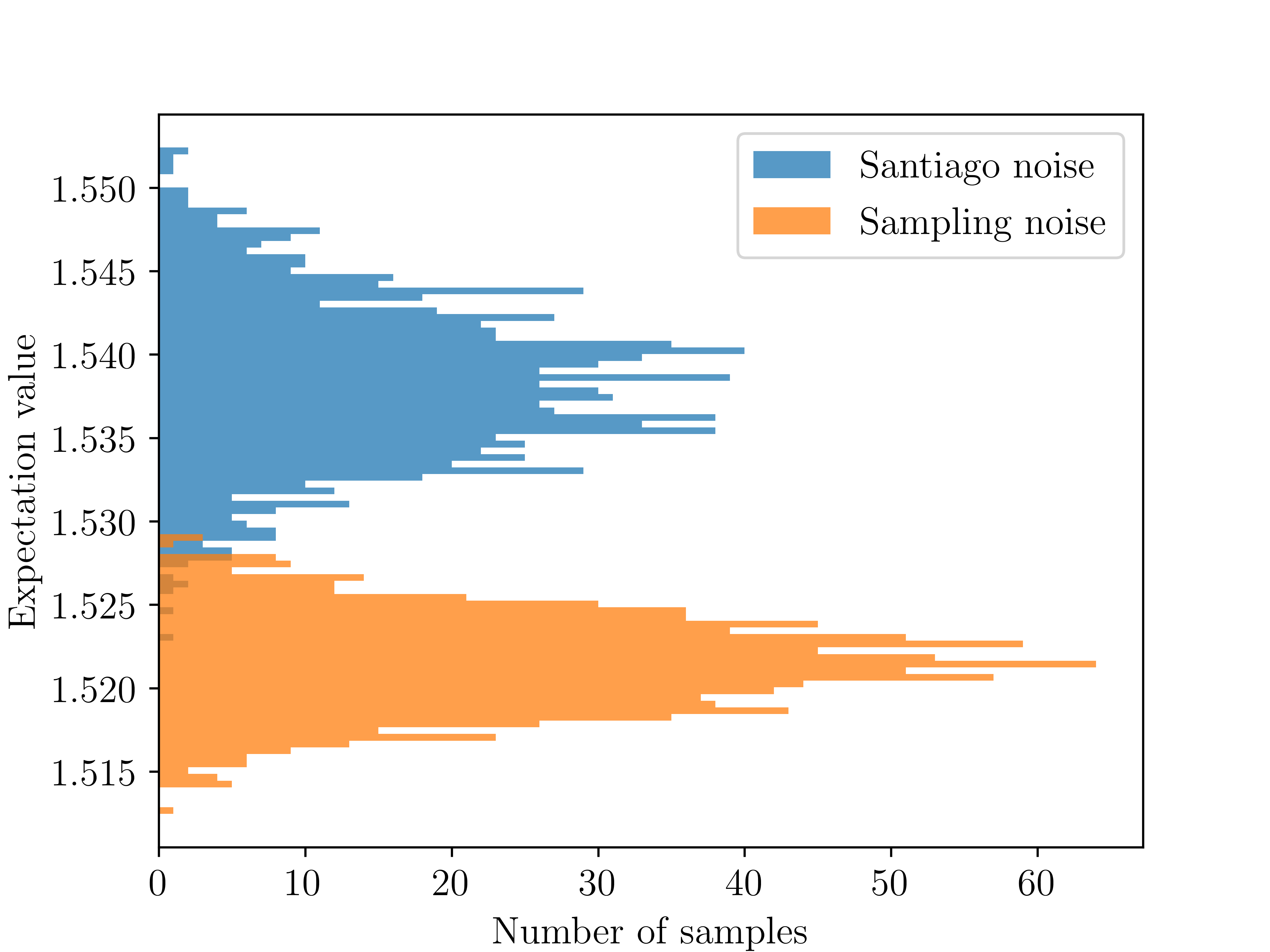}
    \caption{Expectation value distribution obtained from $1000$ circuit (each one executed $20000$ times) on the same converged RyRz parameter set for a $2$ qubit system with $\Delta_\text{r}=0.5k_BT$ for the bi-stable dihedral. Blue distribution relates to the sampling under the simulated IBM Q Santiago hardware noise; orange distribution is due to finite measurements.}
    \label{fig:ExpVal_stat}
\end{figure}

At this point we continue our analysis by discussing the effects of different noise sources on the circuits presented above with the aim of answering the question: how does the introduced noise impact the expectation value measurement for a given circuit configuration? This is especially important in the case of nearly converged configurations in which the error introduced by noise can significantly affect the ability of the optimizer to operate. In order to investigate this point, a converged set of RyRz parameters has been used to study the effects of finite measurements and quantum noise on the expectation value. The optimized variational parameters have been obtained using the IBM Q Santiago quantum computer noise-model. To accumulate statistics, we have measured $\langle \hat{\Tilde{\Gamma}} \rangle$ with $1000$ independent runs. Each Pauli string measurement was accomplished accumulating $20000$ shots. Figure~\ref{fig:ExpVal_stat} reports the results of this investigation as two distinct distributions for the expectation value in the case of a sole sampling noise or with an addition of a noise model tailored on the IBM Q Santiago device.

Looking at the distributions we can easily verify how the introduction of a noise model in the simulation causes both an increased distribution amplitude, with the loss of the peaked Gaussian profile observed for the noiseless simulator, and a shift of the average value. This last point is quite interesting since it highlights the profound difference existing between the two effects. The sampling effect stems from an inherent rounding error due to the limited number of measurement performed. As such, for a set of repeated measures it will appear as a Gaussian distribution centered around the exact expectation value for the state prepared by the variational form. Moreover, the variance of the sampling distribution tends to zero as the number of measurements increases. On the other hand, the simulated noise of a NISQ device alters the generation of the trial state producing slightly different wavefunctions. Since the trial state, obtained via the optimization procedure, is located near a minimum, most of the error in the state preparation will result in a higher expectation value estimate. The best value obtained in the presence of noise is, indeed, $0.7 \%$ higher than the best value obtained in an ideal calculation.

We now consider how the noise affects the quantum calculation of both different models, by changing the potential energy barrier of the conformational transition, and of different implementations, by changing the number of basis functions describing the target eigenfunction (i.e. using a different number of qubits). 

\begin{figure}[b!]
    \centering
    \includegraphics[width=\columnwidth]{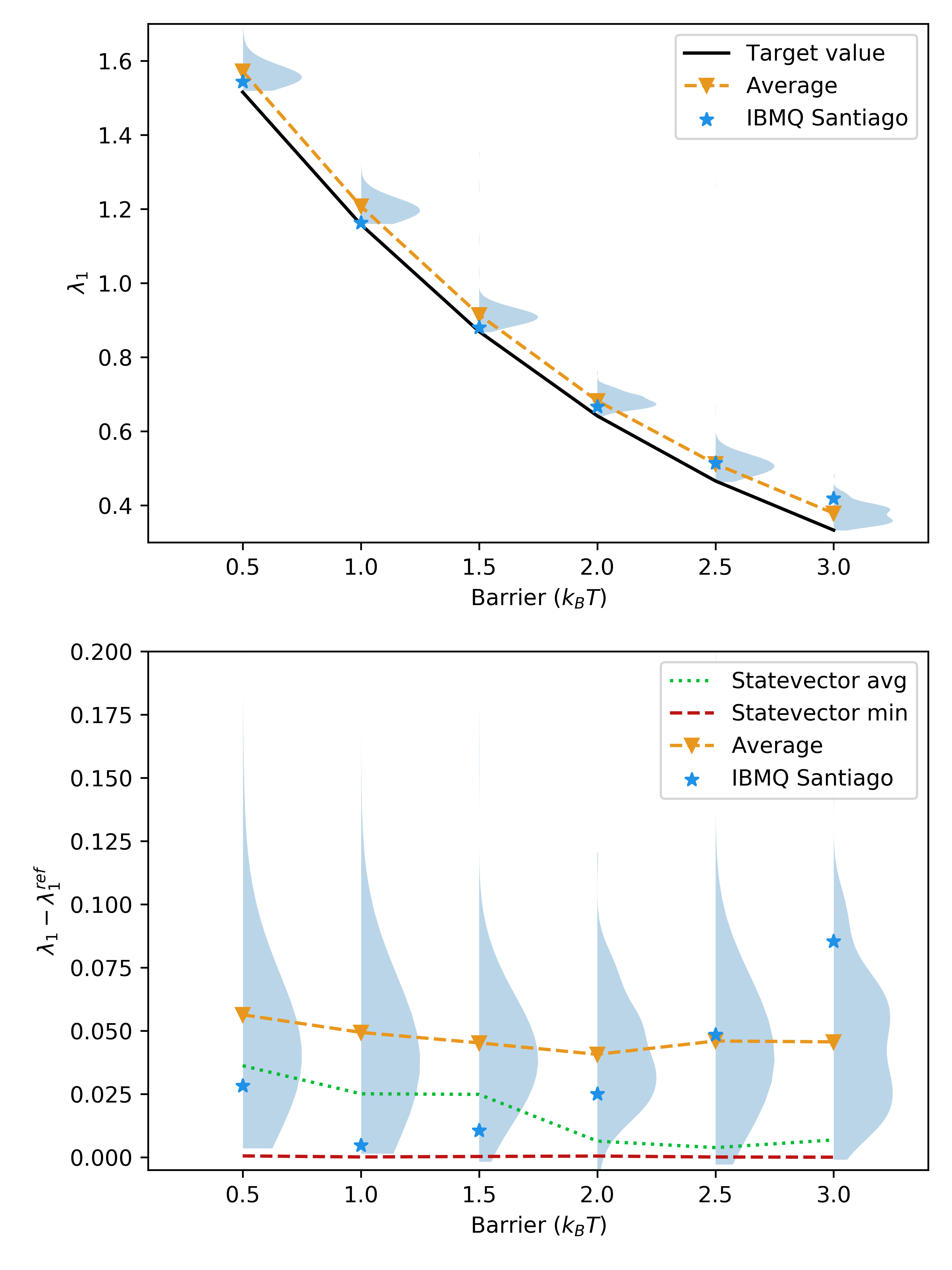}
    \caption{(Upper panel) - Dependence of the first non-zero eigenvalue $\lambda_1$ with respect to the dihedral barrier height $\Delta_\text{r}$. In light blue we report the distribution obtained from $250$ VQE optimizations with random initial guess simulating the circuit execution with a noise model based on the IBMQ Santiago hardware. The black line represents the target value computed for the given basis set while the orange dashed line connects the average value for each group of VQE runs. Blue stars show the results obtained evaluating the last variational circuit (already optimized by the Statevector simulator) on the 5 qubits chip IBMQ Santiago. 
    (Lower panel) - Absolute deviation from the target values. Dotted green line shows the average result obtained with a Statevector simulation repeated 60 times with different initial guesses.}
    \label{fig:2Q_VQE}
\end{figure}

\paragraph{Performance dependency on the potential barrier height}
\hfill
\\
For this evaluation we have performed $250$ VQE routines starting with a random guess for the RyRz parameters. The variational parameters have been optimized using the SPSA algorithm under the effect of the Santiago quantum computer noise model. Every circuit has been transpiled before the simulation.
In Fig. \ref{fig:2Q_VQE} we report the results where two qubits encode the composite dihedral states describing a system of three rotors in which $\Delta_\text{r}$ for the double minimum dihedral has been varied form $0.5 \, k_BT$ to $3.0 \, k_BT$ while $\Delta_\text{nr}$ for the single minimum dihedral was kept fixed at $1 \, k_BT$. 

First of all we can notice how, as expected for activated processes, the kinetic constant decreases as we increase the potential energy barrier. Looking at the data we can observe the good agreement between the classically estimated target value (solid line) and the average results obtained with IBMQ Santiago noise model (downward triangles). In the same graph, we report (stars) the values of $\lambda_1$ obtained computing the expectation value on the IBMQ Santiago quantum computer starting from a set of variational parameters optimized using a VQE procedure carried out on the Qiskit Statevector Simulator. During the expectation value measurement on the IBMQ quantum hardware each circuit has been repeated, in order to build the required statistics, for the maximum allowed number of times (8192 runs). The obtained data, despite being computed from parameters obtained in a completely noiseless procedure, provide a natural metric to look at the results in term of the characteristic noise of a NISQ era quantum computer. These information, together with the ones coming from the noisy simulations, can be seen as a general proof of concept showing the feasibility to run the entire calculation on a NISQ device.\\

Bearing in mind that the quantum noise affects only the evaluation of the final expectation value we may notice that adding a further noise source causes sensible deviations from the noise-free average. To comment this results we leverage the intuition built from the statistical analysis in the previous section: the points collected on the IBM Q Santiago can be seen as samples of a third distribution generated by the noise of the actual quantum hardware. Under this light we may notice that in almost all cases we obtain absolute errors comparable with the maximum and mean deviations of the two distributions reported in Fig. 4. 

This result on the one hand supports the effectiveness of the noise model in capturing the performance of real hardware, and on the other hand allows us to justify the greater deviation from the ideal average at the last point.

Finally, we can also notice that the absolute width of the distribution remains substantially unchanged in the range of barriers explored. As a consequence, the relative error decreases moving from higher to smaller barriers passing from $13.6\%$ at $3.0 \, k_BT$ to $3.72\%$ at $0.5 \, k_BT$. In the same figure (lower panel) the noise-free statevector results are reported as reference. Here we can appreciate with more the detail that, even though the distribution width is almost constant, the optimization is more sensitive with respect to the initial guess distribution for small barrier values. 

\begin{figure}[t!]
    \centering
    \includegraphics[width=\columnwidth]{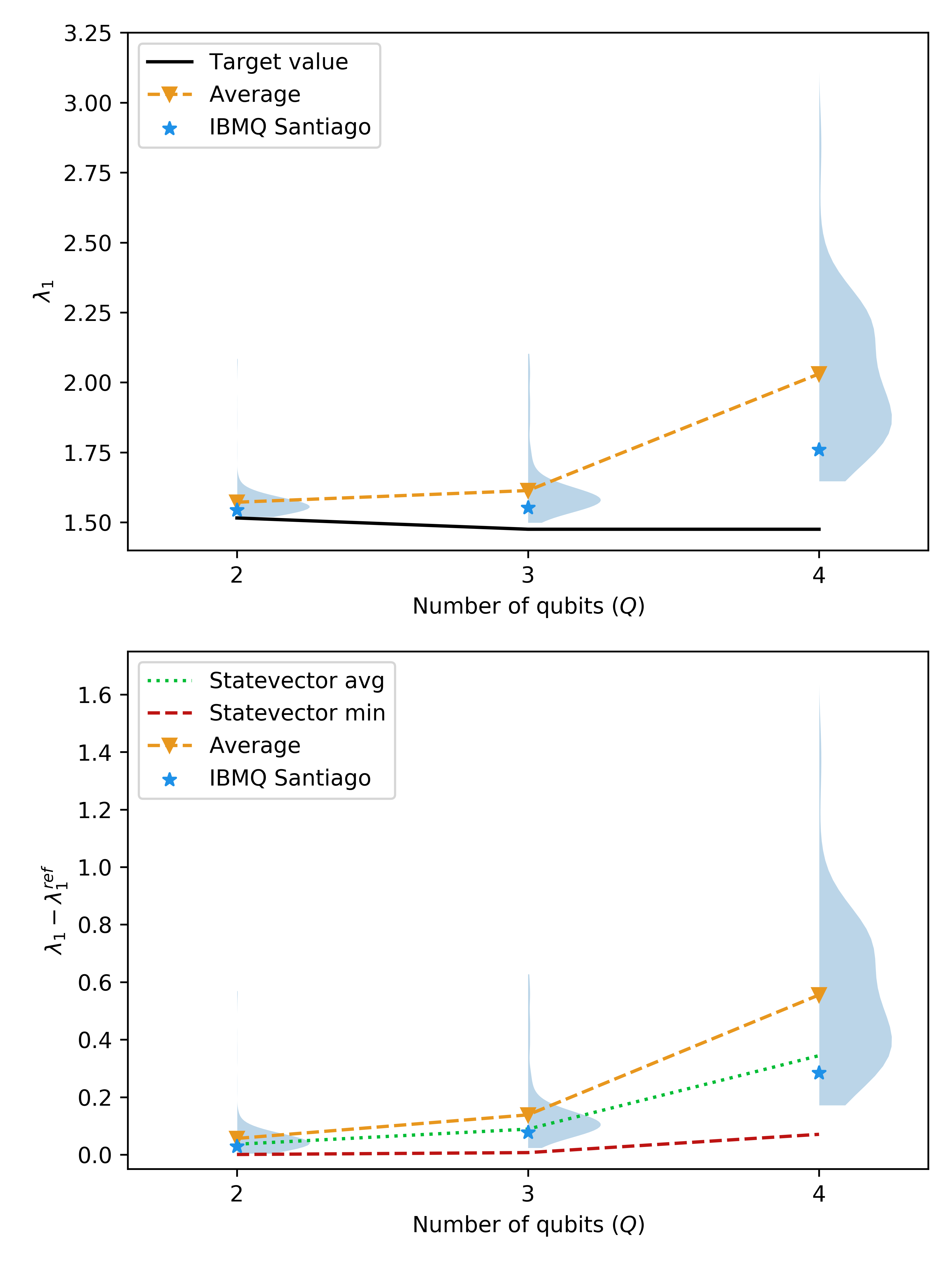}
    \caption{ (Upper panel) - Dependence of the first non-zero eigenvalue $\lambda_1$ with respect to the number of qubits adopted. In  light blue we report the distribution obtained from $250$ VQE optimizations with random initial guess with a noise model based on the IBMQ Santiago hardware. The black line represents the target value computed for the given basis set while the orange dashed line connects the average value for each group of VQE runs. Blue stars show the results obtained evaluating the last variational circuit (already optimized by the Statevector simulator) on the 5 qubits chip IBMQ Santiago. 
    (Lower panel) - Absolute deviation from the target values. Dotted green line shows the average result obtained with a Statevector simulation repeated 60 times with different initial guesses.}
    \label{fig:nQ_VQE}
\end{figure}

\paragraph{Performance dependency on the basis set size}
\hfill
\\
Now that the accuracy of the algorithm has been discussed in terms of the physics of the simulated system, a final question still remains open: how does the performance of the protocol scales with the number of qubits employed? To answer to such a question we consider the same system of a three rotor chain with reactive barrier height of $\Delta_\text{r}=0.5k_BT$. The probability density distribution has been described adopting an increasing number of basis elements from 4 up to 16 odd composite functions (see Eq. \ref{composite_basis_set}). This translates to an increasing number of qubits to encode the overall probability density distribution. The purpose of this analysis is not to assess the accuracy of the basis set size (which determines only a slight change in the kinetic rate constant) but to observe the effect of a noisy device upon performing the same calculation with an increasing number of quantum resources. The results obtained in such a simulation, together with the reference values, are reported in Fig.~\ref{fig:nQ_VQE}.\\ 

As we can see the distribution width rapidly increases with the number of qubits and so does also the average expectation value computed with both statevector and noisy simulators. This points out the increased complexity of the optimization procedure as previously mentioned in Sec. \ref{subsec:variational_network}. For a $4$ qubits system an error of the $37.6\%$ is recovered for the noisy simulation while an error of the $23.3\%$ is encountered in the case of a Statevector simulation. These data confirm the role of the noise in creating less accurate predictions but shows how a significant part of the error itself should be ascribed to a difficult convergence of the classical optimization as anticipated in section~\ref{subsec:variational_network} . This massive performance degradation has also the effect of hiding the tiny margin for accuracy improvement induced by the basis-set extension. This is a critical aspect to consider in this early VQE implementations in which a good trade-off between algorithm stability and basis-set extension must be found. Further analyses of possible alternative variational networks could circumvent this issue.

\section{Conclusions}\label{sec:Conclusions}

In this paper we have presented a novel application for quantum computers. Particularly, we have addressed the solution of the Fokker-Planck-Smoluchowski eigenvalue problem exploiting its isomorphism with the Schrodinger equation. 

The proposed algorithm consists of three main ingredients: (i) a strategy to map the classical probability distribution onto the quantum register, (ii) the choice of a unitary ansatz to generate trial probability distributions (in the same spirit of the original VQEs for quantum systems) and (iii) the implementation of a classical optimization routine updating the unitary control parameters to move towards the exact configuration. Further investigation should follow along all the lines mentioned above.

Indeed, in order to devise a quantum algorithm useful for everyday applications and able to achieve quantum advantage with respect to a classical solution, a polynomial scaling of the quantum resources with the system size is desirable. Our analysis in Sec. \ref{sec:implementation} has shown how this has already partially been achieved considering the information storage point of view. On the other hand, a dense mapping such as the binary approach adopted insofar requires a number of measurements that scales exponentially with the number of rotors in the chain. This aspect can be mitigated by implementing different strategies that group sets of Pauli strings reducing the overall measurement count~\cite{hamamura2020efficient, huggins2021efficient, izmaylov2019unitary, bravyi2017tapering}. For instance, considering the $J$ diagonal elements, a single circuit execution is sufficient since all the strings are diagonal in the computational basis and can be measured simultaneously in a single post-rotation. Despite this, looking at systems of interest from an application point of view a step forward is still needed. 

In a future work we will seek to overcome this issue by developing a mapping on the single-dihedral states which should reduce dramatically the number of terms in the FPS operator representation. In quantum chemistry parlance, this would correspond to a mapping on the spin-orbitals in place of a mapping on the Slater determinant basis.

Other matters to be addressed concern the unitary ansatz generating the trial probability distribution and the classical optimizer. These aspects have been considered in Sec. \ref{sec:Results}. There, we have shown that adopting a RyRz heuristic variational network allows to achieve good results for a small basis set. We believe that the results we have obtained in presence of a simulated noise can be further improved by the incorporation of a mitigation routine such as the zero noise error extrapolation~\cite{giurgica2020digital}.
However, we decided to not include also this feature in the implementation as it goes beyond the purpose of the present study of presenting a new application of hybrid algorithms in the field of stochastic processes.

At the same time, we have also noticed that the performance scales poorly with respect to the number of qubits and suffers of some problems deriving from the uniform and agnostic coverage of the underlying Hilbert space. In this direction we have suggested as a working strategy to lower the barrier of the optimization task by applying an hierarchical procedure in which the solution of the variational optimization with a smaller number of basis function is used as initial guess for the estimation of the kinetic rate constant using an higher number of qubits. Such an approach is in line with the development of algorithms in the NISQ era where small quantum processors are put to test by very shallow circuits; alternatively, we should look at further refinements of this implementation by developing physically-inspired ansatze as done in other contexts. Also this aspect will be investigated in a future work.
 
In essence, we achieved the proof of concept of the application of a quantum computer to solve stochastic dynamics. With some technical developments as discussed in the previous paragraphs, the proposed procedure brings stochastic dynamics amongst the promising applications of future quantum devices.

\section*{Acknowledgements}

We are grateful to Prof. Ivano Tavernelli for fruitful discussions. We acknowledge the use of IBM Quantum services for this work. The views expressed are those of the authors, and do not reflect the official policy or position of IBM or the IBM Quantum team. We acknowledge the support by the DOR funding schemes of the Dipartimento di Scienze Chimiche, Università di Padova. P.P. is grateful to Fondazione CARIPARO for financial support (PhD grant). P.P. is grateful to the Cloud Veneto computing infrastructure for the computation time granted under the project "Nuclear quantum effects". D.C. is grateful to MIUR "Dipartimenti di Eccellenza" under the project Nanochemistry for energy and Health (NExuS) for funding the PhD grant.

\setcounter{section}{0}
\setcounter{table}{0}
\def\thesection{\Alph{section}}
\def\thesubsection{A\arabic{section}.\arabic{subsection}}
\def\thesubsubsection{A\arabic{section}.\arabic{subsection}.\arabic{subsubsection}}
\def\thetable{A\arabic{table}} 

\section*{Appendix A}

\begin{figure}[h!]
    \centering
    \includegraphics[width=\columnwidth]{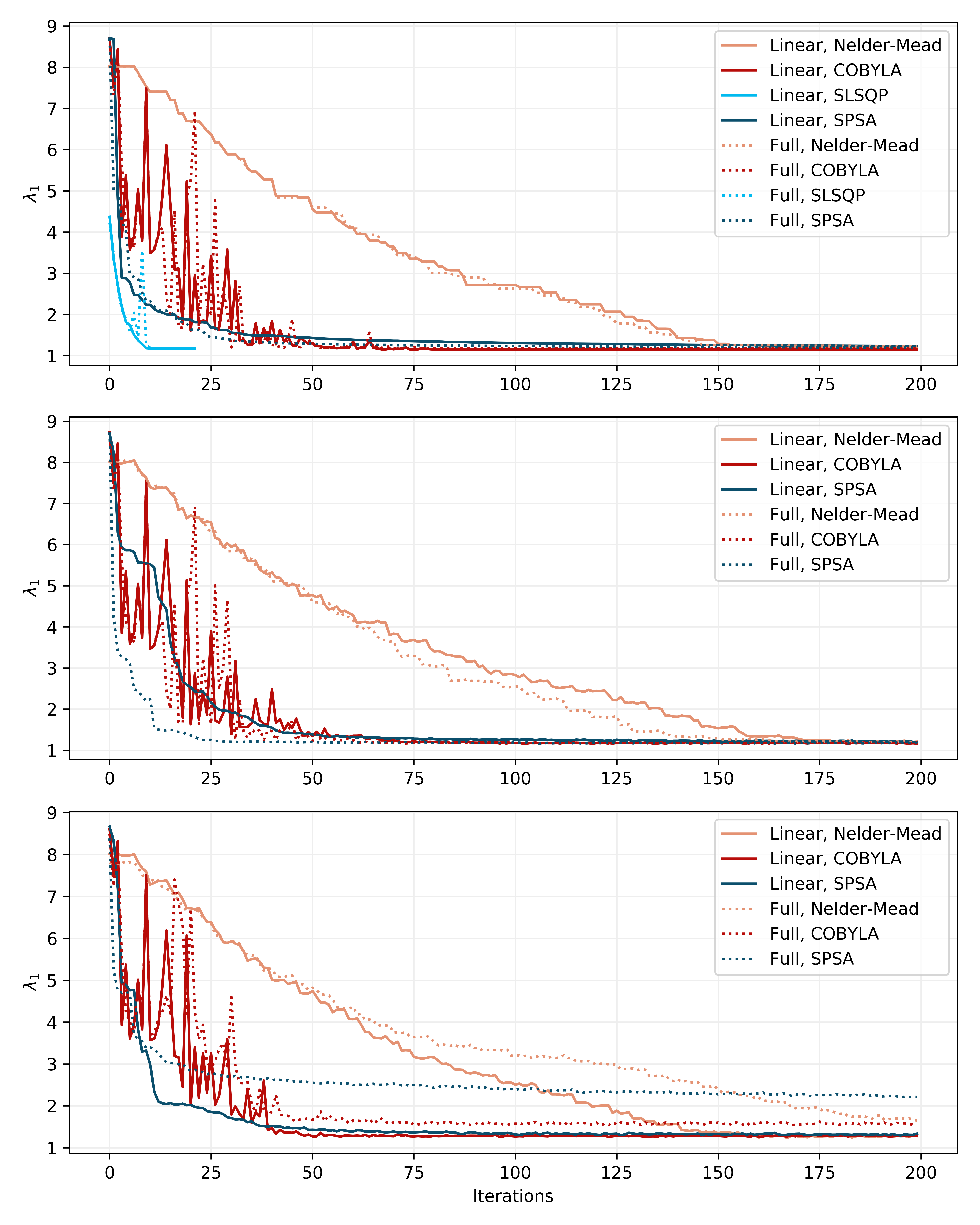}
    \caption{Comparison between single VQE optimization under different conditions of optimizer and entangler. The tree graphs refers to the results obtained with the Qiskit Statevector Simulator (higher panel), Qiskit QASM without (central panel) and with (lower panel) the IBMQ Santiago noise model. For the middle and lower panel the number of measures on the same circuit has been set to 20000.}
    \label{fig:single_VQE_data}
\end{figure}

In what follows the optimizer selection and the effect that the entangler type has on the convergence behavior of the algorithm will be presented. Addressing the discussion of this problem is not trivial since the convergence profile of each VQE routine depends upon the selected set of initial parameters. In what follows we will proceed by analyzing, for different choices of optimizer and entangler, the convergence profile of a single VQE. This clearly represents only a single point in the multidimensional initial parameter space and consequently only comparative considerations can be done between different methods. Such a simple data representation, visible in Figure~\ref{fig:single_VQE_data}, can anyway already give a useful qualitative picture. For the present tests, a $3$ qubit system is considered; both dihedral barriers (single and double) have been set to $1\, k_B\, T$ and a common set of randomly selected initial parameters has been kept fixed for all simulations. The built-in Qiskit SPSA optimizer together with the Nelder-Mead, SLSQP and COBYLA optimizer from the SciPy~\cite{scipy2020} Python module have been employed. The figure contains three plots that represent the profile of the VQE optimization in the case of various type of noise. The top panel contains data obtained using the noiseless statevector simulator. The center panel represents a noiseless simulator in which the number of measures on the same circuit has been set to $20000$. The lower panel shows the results for the transpiled circuits simulated including the noise model of the IBMQ Santiago quantum computer. Also for this last case the number of runs has been set to $20000$.\\

Analyzing Figure~\ref{fig:single_VQE_data} many interesting comments can be made. First of all looking at the raw performance of the various optimizers one can easily observe how the Nelder-Mead algorithm consistently shows a slower convergence toward the target value if compared with the other optimization procedures. The SLSQP algorithm shows very good performance in the Statevector simulations converging in few iterations to the target value; however its sensitivity to noise prevent, as far as we can tell, its application to noisy simulations. Both COBYLA and SPSA shows consistently stable and reasonably good performances with the latter showing a more consistent behaviour and greater accuracy in simulations with higher noise. For this reason SPSA has been used as the default optimizer throughout the paper. Looking at the central panel one can easily see how little difference is observed between different entanglers in term of converged values with the full entangler resulting mildly more accurate. As it can be seen in the lower panel, the situation is rapidly reversed by the introduction of a realistic quantum computer noise model. The more complex structure of the full entangler joined with the less efficient transpilation, results in a slower converging and significantly less accurate result, clearly pointing to the linear entangler as the natural candidate for a realistic implementation.\\

\printbibliography

@article{Peruzzo2014,
author = {Peruzzo, Alberto and McClean, Jarrod and Shadbolt, Peter and Yung, Man Hong and Zhou, Xiao Qi and Love, Peter J. and Aspuru-Guzik, Al{\'{a}}n and O'Brien, Jeremy L.},
doi = {10.1038/ncomms5213},
issn = {20411723},
journal = {Nature Communications},
month = {7},
number = {1},
pages = {1--7},
publisher = {Nature Publishing Group},
title = {{A variational eigenvalue solver on a photonic quantum processor}},
url = {www.nature.com/naturecommunications},
volume = {5},
year = {2014}
}

@article{McArdle2020,
archivePrefix = {arXiv},
arxivId = {1808.10402},
author = {McArdle, Sam and Endo, Suguru and Aspuru-Guzik, Al{\'{a}}n and Benjamin, Simon C. and Yuan, Xiao},
doi = {10.1103/RevModPhys.92.015003},
eprint = {1808.10402},
issn = {15390756},
journal = {Reviews of Modern Physics},
month = {mar},
number = {1},
pages = {015003},
publisher = {American Physical Society},
title = {{Quantum computational chemistry}},
url = {https://journals.aps.org/rmp/abstract/10.1103/RevModPhys.92.015003},
volume = {92},
year = {2020}
}

@article{Kandala2017,
author = {Kandala, Abhinav and Mezzacapo, Antonio and Temme, Kristan and maika Takita and markus Brink and Chow, Jerry and Gambetta, Jay},
doi = {10.1038/nature23879},
title = {{Hardware-efficient variational quantum eigensolver for small molecules and quantum magnets}},
year = {2017}
}

@article{Ollitrault2020,
archivePrefix = {arXiv},
arxivId = {2003.12578},
author = {Ollitrault, Pauline J. and Baiardi, Alberto and Reiher, Markus and Tavernelli, Ivano},
doi = {10.1039/d0sc01908a},
eprint = {2003.12578},
issn = {20416539},
journal = {Chemical Science},
month = {jul},
number = {26},
pages = {6842--6855},
publisher = {Royal Society of Chemistry},
title = {{Hardware efficient quantum algorithms for vibrational structure calculations}},
url = {https://pubs.rsc.org/en/content/articlehtml/2020/sc/d0sc01908a https://pubs.rsc.org/en/content/articlelanding/2020/sc/d0sc01908a},
volume = {11},
year = {2020}
}

@article{Kramers1940,
author = {Kramers, H. A.},
doi = {10.1016/S0031-8914(40)90098-2},
issn = {00318914},
journal = {Physica},
month = {apr},
number = {4},
pages = {284--304},
publisher = {North-Holland},
title = {{Brownian motion in a field of force and the diffusion model of chemical reactions}},
volume = {7},
year = {1940}
}

@article{Aleksandrowicz2019,
author = {Aleksandrowicz, Gadi and Alexander, Thomas and Barkoutsos, Panagiotis and Bello, Luciano and Ben-Haim, Yael and Bucher, David and Cabrera-Hern{\'{a}}ndez, Francisco Jose and Carballo-Franquis, Jorge and Chen, Adrian and Chen, Chun-Fu and Chow, Jerry M. and C{\'{o}}rcoles-Gonzales, Antonio D. and Cross, Abigail J. and Cross, Andrew and Cruz-Benito, Juan and Culver, Chris and Gonz{\'{a}}lez, Salvador De La Puente and Torre, Enrique De La and Ding, Delton and Dumitrescu, Eugene and Duran, Ivan and Eendebak, Pieter and Everitt, Mark and Sertage, Ismael Faro and Frisch, Albert and Fuhrer, Andreas and Gambetta, Jay and Gago, Borja Godoy and Gomez-Mosquera, Juan and Greenberg, Donny and Hamamura, Ikko and Havlicek, Vojtech and Hellmers, Joe and Herok, {\L}ukasz and Horii, Hiroshi and Hu, Shaohan and Imamichi, Takashi and Itoko, Toshinari and Javadi-Abhari, Ali and Kanazawa, Naoki and Karazeev, Anton and Krsulich, Kevin and Liu, Peng and Luh, Yang and Maeng, Yunho and Marques, Manoel and Mart{\'{i}}n-Fern{\'{a}}ndez, Francisco Jose and McClure, Douglas T. and McKay, David and Meesala, Srujan and Mezzacapo, Antonio and Moll, Nikolaj and Rodr{\'{i}}guez, Diego Moreda and Nannicini, Giacomo and Nation, Paul and Ollitrault, Pauline and O'Riordan, Lee James and Paik, Hanhee and P{\'{e}}rez, Jes{\'{u}}s and Phan, Anna and Pistoia, Marco and Prutyanov, Viktor and Reuter, Max and Rice, Julia and Davila, Abd{\'{o}}n Rodr{\'{i}}guez and Rudy, Raymond Harry Putra and Ryu, Mingi and Sathaye, Ninad and Schnabel, Chris and Schoute, Eddie and Setia, Kanav and Shi, Yunong and Silva, Adenilton and Siraichi, Yukio and Sivarajah, Seyon and Smolin, John A. and Soeken, Mathias and Takahashi, Hitomi and Tavernelli, Ivano and Taylor, Charles and Taylour, Pete and Trabing, Kenso and Treinish, Matthew and Turner, Wes and Vogt-Lee, Desiree and Vuillot, Christophe and Wildstrom, Jonathan A. and Wilson, Jessica and Winston, Erick and Wood, Christopher and Wood, Stephen and W{\"{o}}rner, Stefan and Akhalwaya, Ismail Yunus and Zoufal, Christa},
doi = {10.5281/ZENODO.2562111},
month = {jan},
title = {{Qiskit: An Open-source Framework for Quantum Computing}},
url = {https://zenodo.org/record/2562111},
year = {2019}
}

@article{scipy2020,
  author  = {Virtanen, Pauli and Gommers, Ralf and Oliphant, Travis E. and Haberland, Matt and Reddy, Tyler and Cournapeau, David and Burovski, Evgeni and Peterson, Pearu and Weckesser, Warren and Bright, Jonathan and {van der Walt}, St{\'e}fan J. and Brett, Matthew and Wilson, Joshua and Millman, K. Jarrod and Mayorov, Nikolay and Nelson, Andrew R. J. and Jones, Eric and Kern, Robert and Larson, Eric and Carey, C J and Polat, {\.I}lhan and Feng, Yu and Moore, Eric W. and {VanderPlas}, Jake and Laxalde, Denis and Perktold, Josef and Cimrman, Robert and Henriksen, Ian and Quintero, E. A. and Harris, Charles R. and Archibald, Anne M. and Ribeiro, Ant{\^o}nio H. and Pedregosa, Fabian and {van Mulbregt}, Paul and {SciPy 1.0 Contributors}},
  title   = {{{SciPy} 1.0: Fundamental Algorithms for Scientific  Computing in Python}},
  journal = {Nature Methods},
  year    = {2020},
  volume  = {17},
  pages   = {261--272},
  adsurl  = {https://rdcu.be/b08Wh},
  doi     = {10.1038/s41592-019-0686-2},
}

@article{deutsch2020harnessing,
  title={Harnessing the power of the second quantum revolution},
  author={Deutsch, Ivan H},
  journal={PRX Quantum},
  volume={1},
  number={2},
  pages={020101},
  year={2020},
  publisher={APS}
}

@article{terhal2018quantum,
  title={Quantum supremacy, here we come},
  author={Terhal, Barbara M},
  journal={Nature Physics},
  volume={14},
  number={6},
  pages={530--531},
  year={2018},
  publisher={Nature Publishing Group}
}

@article{georgescu2014quantum,
  title={Quantum simulation},
  author={Georgescu, Iulia M and Ashhab, Sahel and Nori, Franco},
  journal={Reviews of Modern Physics},
  volume={86},
  number={1},
  pages={153},
  year={2014},
  publisher={APS}
}

@article{mcclean2016theory,
  title={The theory of variational hybrid quantum-classical algorithms},
  author={McClean, Jarrod R and Romero, Jonathan and Babbush, Ryan and Aspuru-Guzik, Al{\'a}n},
  journal={New Journal of Physics},
  volume={18},
  number={2},
  pages={023023},
  year={2016},
  publisher={IOP Publishing}
}

@article{jurcevic2021demonstration,
  title={Demonstration of quantum volume 64 on a superconducting quantum computing system},
  author={Jurcevic, Petar and Javadi-Abhari, Ali and Bishop, Lev S and Lauer, Isaac and Borgorin, Daniela and Brink, Markus and Capelluto, Lauren and Gunluk, Oktay and Itoko, Toshinari and Kanazawa, Naoki and others},
  journal={Quantum Science and Technology},
  year={2021},
  publisher={IOP Publishing}
}

@article{yoshioka2020variational,
  title={Variational Quantum Simulation for Periodic Materials},
  author={Yoshioka, Nobuyuki and Mizukami, Wataru},
  journal={arXiv preprint arXiv:2008.09492},
  year={2020}
}

@article{nachman2021quantum,
  title={Quantum Algorithm for High Energy Physics Simulations},
  author={Nachman, Benjamin and Provasoli, Davide and de Jong, Wibe A and Bauer, Christian W},
  journal={Physical Review Letters},
  volume={126},
  number={6},
  pages={062001},
  year={2021},
  publisher={APS}
}

@article{haven2002discussion,
  title={A discussion on embedding the Black--Scholes option pricing model in a quantum physics setting},
  author={Haven, Emmanuel E},
  journal={Physica A: Statistical Mechanics and its Applications},
  volume={304},
  number={3-4},
  pages={507--524},
  year={2002},
  publisher={Elsevier}
}

@article{lin2014long,
  title={Long-time behavior of a stochastic SIR model},
  author={Lin, Yuguo and Jiang, Daqing and Xia, Peiyan},
  journal={Applied Mathematics and Computation},
  volume={236},
  pages={1--9},
  year={2014},
  publisher={Elsevier}
}

@article{singh2016gaseous,
  title={Gaseous microflow modeling using the Fokker-Planck equation},
  author={Singh, SK and Thantanapally, Chakradhar and Ansumali, Santosh},
  journal={Physical Review E},
  volume={94},
  number={6},
  pages={063307},
  year={2016},
  publisher={APS}
}

@article{colless2018computation,
  title={Computation of molecular spectra on a quantum processor with an error-resilient algorithm},
  author={Colless, James I and Ramasesh, Vinay V and Dahlen, Dar and Blok, Machiel S and Kimchi-Schwartz, ME and McClean, JR and Carter, J and De Jong, WA and Siddiqi, I},
  journal={Physical Review X},
  volume={8},
  number={1},
  pages={011021},
  year={2018},
  publisher={APS}
}

@article{ollitrault2020quantum,
  title={Quantum equation of motion for computing molecular excitation energies on a noisy quantum processor},
  author={Ollitrault, Pauline J and Kandala, Abhinav and Chen, Chun-Fu and Barkoutsos, Panagiotis Kl and Mezzacapo, Antonio and Pistoia, Marco and Sheldon, Sarah and Woerner, Stefan and Gambetta, Jay M and Tavernelli, Ivano},
  journal={Physical Review Research},
  volume={2},
  number={4},
  pages={043140},
  year={2020},
  publisher={APS}
}

@article{higgott2019variational,
  title={Variational quantum computation of excited states},
  author={Higgott, Oscar and Wang, Daochen and Brierley, Stephen},
  journal={Quantum},
  volume={3},
  pages={156},
  year={2019},
  publisher={Verein zur F{\"o}rderung des Open Access Publizierens in den Quantenwissenschaften}
}

@article{hamamura2020efficient,
  title={Efficient evaluation of quantum observables using entangled measurements},
  author={Hamamura, Ikko and Imamichi, Takashi},
  journal={npj Quantum Information},
  volume={6},
  number={1},
  pages={1--8},
  year={2020},
  publisher={Nature Publishing Group}
}

@article{izmaylov2019unitary,
  title={Unitary partitioning approach to the measurement problem in the variational quantum eigensolver method},
  author={Izmaylov, Artur F and Yen, Tzu-Ching and Lang, Robert A and Verteletskyi, Vladyslav},
  journal={Journal of chemical theory and computation},
  volume={16},
  number={1},
  pages={190--195},
  year={2019},
  publisher={ACS Publications}
}

@article{huggins2021efficient,
  title={Efficient and noise resilient measurements for quantum chemistry on near-term quantum computers},
  author={Huggins, William J and McClean, Jarrod R and Rubin, Nicholas C and Jiang, Zhang and Wiebe, Nathan and Whaley, K Birgitta and Babbush, Ryan},
  journal={npj Quantum Information},
  volume={7},
  number={1},
  pages={1--9},
  year={2021},
  publisher={Nature Publishing Group}
}

@article{bravyi2017tapering,
  title={Tapering off qubits to simulate fermionic Hamiltonians},
  author={Bravyi, Sergey and Gambetta, Jay M and Mezzacapo, Antonio and Temme, Kristan},
  journal={arXiv preprint arXiv:1701.08213},
  year={2017}
}

@article{moll2018quantum,
  title={Quantum optimization using variational algorithms on near-term quantum devices},
  author={Moll, Nikolaj and Barkoutsos, Panagiotis and Bishop, Lev S and Chow, Jerry M and Cross, Andrew and Egger, Daniel J and Filipp, Stefan and Fuhrer, Andreas and Gambetta, Jay M and Ganzhorn, Marc and others},
  journal={Quantum Science and Technology},
  volume={3},
  number={3},
  pages={030503},
  year={2018},
  publisher={IOP Publishing}
}

@article{sim2019expressibility,
  title={Expressibility and Entangling Capability of Parameterized Quantum Circuits for Hybrid Quantum-Classical Algorithms},
  author={Sim, Sukin and Johnson, Peter D and Aspuru-Guzik, Alainn},
  journal={Advanced Quantum Technologies},
  volume={2},
  number={12},
  pages={1900070},
  year={2019},
  publisher={Wiley Online Library}
}

@article{wecker2015progress,
  title={Progress towards practical quantum variational algorithms},
  author={Wecker, Dave and Hastings, Matthew B and Troyer, Matthias},
  journal={Physical Review A},
  volume={92},
  number={4},
  pages={042303},
  year={2015},
  publisher={APS}
}

@article{babbush2018low,
  title={Low-depth quantum simulation of materials},
  author={Babbush, Ryan and Wiebe, Nathan and McClean, Jarrod and McClain, James and Neven, Hartmut and Chan, Garnet Kin-Lic},
  journal={Physical Review X},
  volume={8},
  number={1},
  pages={011044},
  year={2018},
  publisher={APS}
}

@article{woitzik2020entanglement,
  title={Entanglement production and convergence properties of the variational quantum eigensolver},
  author={Woitzik, Andreas JC and Barkoutsos, Panagiotis Kl and Wudarski, Filip and Buchleitner, Andreas and Tavernelli, Ivano},
  journal={Physical Review A},
  volume={102},
  number={4},
  pages={042402},
  year={2020},
  publisher={APS}
}

@article{popescu2006entanglement,
  title={Entanglement and the foundations of statistical mechanics},
  author={Popescu, Sandu and Short, Anthony J and Winter, Andreas},
  journal={Nature Physics},
  volume={2},
  number={11},
  pages={754--758},
  year={2006},
  publisher={Nature Publishing Group}
}

@article{fresch2013typical,
  title={Typical response of quantum pure states},
  author={Fresch, Barbara and Moro, Giorgio J},
  journal={The European Physical Journal B},
  volume={86},
  number={5},
  pages={1--13},
  year={2013},
  publisher={Springer}
}

@article{mcclean2018barren,
  title={Barren plateaus in quantum neural network training landscapes},
  author={McClean, Jarrod R and Boixo, Sergio and Smelyanskiy, Vadim N and Babbush, Ryan and Neven, Hartmut},
  journal={Nature communications},
  volume={9},
  number={1},
  pages={1--6},
  year={2018},
  publisher={Nature Publishing Group}
}

@article{cerezo2021cost,
  title={Cost function dependent barren plateaus in shallow parametrized quantum circuits},
  author={Cerezo, M and Sone, Akira and Volkoff, Tyler and Cincio, Lukasz and Coles, Patrick J},
  journal={Nature Communications},
  volume={12},
  number={1},
  pages={1--12},
  year={2021},
  publisher={Nature Publishing Group}
}

@book{van1992stochastic,
  title={Stochastic processes in physics and chemistry},
  author={Van Kampen, Nicolaas Godfried},
  volume={1},
  year={1992},
  publisher={Elsevier}
}

@book{gardiner1985handbook,
  title={Handbook of stochastic methods},
  author={Gardiner, Crispin W and others},
  volume={3},
  year={1985},
  publisher={springer Berlin}
}

@book{elber2020molecular,
  title={Molecular kinetics in condensed phases: Theory, simulation, and analysis},
  author={Elber, Ron and Makarov, Dmitrii E and Orland, Henri},
  year={2020},
  publisher={John Wiley \& Sons}
}

@article{miyazawa1989theory,
  title={Theory of the one-variable Fokker-Planck equation},
  author={Miyazawa, Toru},
  journal={Physical Review A},
  volume={39},
  number={3},
  pages={1447},
  year={1989},
  publisher={APS}
}

@article{helfand1978brownian,
  title={Brownian dynamics study of transitions in a polymer chain of bistable oscillators},
  author={Helfand, Eugene},
  journal={The Journal of Chemical Physics},
  volume={69},
  number={3},
  pages={1010--1018},
  year={1978},
  publisher={American Institute of Physics}
}

@article{moro1991coupling,
  title={The coupling between librational motions and conformational transitions in chain molecules. A phenomenological analysis},
  author={Moro, Giorgio J},
  journal={The Journal of chemical physics},
  volume={94},
  number={12},
  pages={8577--8591},
  year={1991},
  publisher={American Institute of Physics}
}

@inproceedings{giurgica2020digital,
  title={Digital zero noise extrapolation for quantum error mitigation},
  author={Giurgica-Tiron, Tudor and Hindy, Yousef and LaRose, Ryan and Mari, Andrea and Zeng, William J},
  booktitle={2020 IEEE International Conference on Quantum Computing and Engineering (QCE)},
  pages={306--316},
  year={2020},
  organization={IEEE}
}

@book{doi1988theory,
  title={The theory of polymer dynamics},
  author={Doi, Masao and Edwards, Samuel Frederick and Edwards, Samuel Frederick},
  volume={73},
  year={1988},
  publisher={oxford university press}
}

@article{moro1992coupling,
  title={The coupling between librational motions and conformational transitions in chain molecules. II. The rotor chain represented by the master equation for site distributions},
  author={Moro, Giorgio J},
  journal={The Journal of chemical physics},
  volume={97},
  number={8},
  pages={5749--5765},
  year={1992},
  publisher={American Institute of Physics}
}

@article{hanggi1990reaction,
  title={Reaction-rate theory: fifty years after Kramers},
  author={H{\"a}nggi, Peter and Talkner, Peter and Borkovec, Michal},
  journal={Reviews of modern physics},
  volume={62},
  number={2},
  pages={251},
  year={1990},
  publisher={APS}
}

@incollection{jungel2016fokker,
  title={Fokker--Planck Equations},
  author={J{\"u}ngel, Ansgar},
  booktitle={Entropy Methods for Diffusive Partial Differential Equations},
  pages={19--44},
  year={2016},
  publisher={Springer}
}

@article{sawaya2020resource,
  title={Resource-efficient digital quantum simulation of d-level systems for photonic, vibrational, and spin-s Hamiltonians},
  author={Sawaya, Nicolas PD and Menke, Tim and Kyaw, Thi Ha and Johri, Sonika and Aspuru-Guzik, Al{\'a}n and Guerreschi, Gian Giacomo},
  journal={npj Quantum Information},
  volume={6},
  number={1},
  pages={1--13},
  year={2020},
  publisher={Nature Publishing Group}
}

@article{castaldo2021quantum,
  title={Quantum optimal control with quantum computers: A hybrid algorithm featuring machine learning optimization},
  author={Castaldo, Davide and Rosa, Marta and Corni, Stefano},
  journal={Physical Review A},
  volume={103},
  number={2},
  pages={022613},
  year={2021},
  publisher={APS}
}

@article{mcardle2019digital,
  title={Digital quantum simulation of molecular vibrations},
  author={McArdle, Sam and Mayorov, Alexander and Shan, Xiao and Benjamin, Simon and Yuan, Xiao},
  journal={Chemical science},
  volume={10},
  number={22},
  pages={5725--5735},
  year={2019},
  publisher={Royal Society of Chemistry}
}

@article{fiuravsek2001maximum,
  title={Maximum-likelihood estimation of quantum measurement},
  author={Fiur{\'a}{\v{s}}ek, Jarom{\'i}r},
  journal={Physical Review A},
  volume={64},
  number={2},
  pages={024102},
  year={2001},
  publisher={APS}
}

@misc{binary_vqe,
 title={},
 author={https://github.com/ppravatto/Binary-VQE}
}

@misc{smoluchowski_rotor_chain,
 title={},
 author={https://github.com/ppravatto/Smoluchowski-Rotor-Chain}
}

\end{document}